# Discovery of Dynamical Heterogeneity in a Supercooled Magnetic Monopole Fluid

Jahnatta Dasini[a,1], Chaia Carroll[a,1], Hiroto Takahashi[b], Jack Murphy[a], Chun-Chih Hsu[b], Sudarshan Sharma[c], Catherine Dawson[a], Fabian Jerzembeck[b,d], Stephen J. Blundell[b], Graeme Luke[c], J.C. Séamus Davis[a,b,d] and Jonathan Ward[a]

**Affiliations:**

a) *Department of Physics, University College Cork, Cork T12 R5C, Ireland*
b) *Clarendon Laboratory, Oxford University, Parks Road, Oxford, OX1 3PU, UK*
c) *McMaster University, Hamilton Ontario, Canada*
d) *Max-Planck Institute for Chemical Physics of Solids, D-01187 Dresden, Germany*

**Dynamical heterogeneity, in which transitory local fluctuations occur in the conformation and dynamics of constituent particles, is widely hypothesized to be essential to evolution of supercooled liquids into the structural glass state. Yet its microscopic spatiotemporal phenomenology is challenging to detect directly in molecular glass forming liquids. Because recent theoretical advances predict that corresponding dynamical heterogeneity could occur in supercooled magnetic monopole fluids (Proc. Nat. Acad. Sci. 112, 8549 (2015)), we searched for such phenomena in $Dy_2Ti_2O_7$. By measuring its microsecond-resolved spontaneous magnetization fluctuations $M(t,T)$ we discovered a sharp bifurcation in monopole noise characteristics below $T \approx 1500$ mK, with the appearance of powerful spontaneous monopole current bursts. This intense dynamics emerges upon entering the supercooled monopole fluid regime, reaches maximum strength near $T \approx 750$ mK and then collapses along with coincident loss of ergodicity approaching $T_g \approx 250$ mK. Moreover, when the four-point dynamical susceptibility $\chi_4(\tau,T)$ is determined directly from temperature dependence of correlations in $M(t,T)$, it evolves as predicted when dynamical heterogeneity is present, revealing its simultaneously and rapidly escalating length and time scales, $\xi(T)$ and $\tau_4(T)$. This overall phenomenology**

**greatly expands our empirical knowledge of supercooled monopole fluids and, more generally, demonstrates direct detection of the time sequence, magnitude, statistics and correlations of dynamical heterogeneity, access to which may greatly accelerate fundamental vitrification studies.**

**Significance Statement:**

Glasses are ubiquitous, yet their microscopic mechanism remains unidentified. A key hypothesis is that supercooled liquids evolve into glasses through spatiotemporal dynamical heterogeneity i.e. transitory local fluctuations in the conformation and dynamics of constituent particles. While four-point dynamic correlation functions could validate such models, these are challenging to detect directly in glass forming liquids. But, because analogous physics may exist for monopoles in spin ice, we searched for dynamical heterogeneity in the supercooled monopole fluid of $Dy_2Ti_2O_7$. Measuring the monopole four-point correlation function directly from magnetization fluctuations, we demonstrate that supercooled monopoles exhibit numerous classic signatures anticipated for dynamical heterogeneity. This highlights both a striking universality in vitrification dynamics and how spin ice offers unique opportunities for dynamical heterogeneity studies.

**1** "*The deepest and most interesting unsolved problem in solid state theory is probably the theory of the nature of glass and the glass transition*" P. W. Anderson (*1*). Although most pure liquids crystallize at their melting temperature, glass-forming liquids instead first enter the supercooled state (*2*, *3*, *4*) and eventually transition into a glass state. During this evolution it is widely hypothesized that the dynamics of constituent particles slow down radically and in an increasingly heterogeneous fashion (*2*-*7*) so that local regions relax on different trajectories at different rates in a continuously evolving yet globally ergodic fashion. These phenomena are thermally activated (*8-13*) events about an unchanging thermodynamic equilibrium. How their atomic-scale phenomenology controls the vitrification process remains an intense focus of modern research (*2-16*). Current theoretical progress includes predictions of frequency-resolved loss of ergodicity (*14*); of trapped nanoscale droplets with internal fluidic particle dynamics (*15*); and of evolution from supercooled dynamical heterogeneity through the glass transition (*16*). Only recently, however, have such phenomena been hypothesized to occur (*17-21*) upon cooling the magnetic monopole fluids of spin-ice.

**2** The most pertinent material is $Dy_2Ti_2O_7$ which contains a sub-lattice of corner-sharing tetrahedra, each having a magnetic $Dy^{3+}$ ion at its four vertices. The Dy magnetic moments ($\mu \approx 10\ \mu_B$) are Ising-like, being constrained to point along their local [111] directions towards or away from the tetrahedron center. The consequent dipolar spin-ice Hamiltonian is (*22*)

$$H = -J\sum_{<ij>} \boldsymbol{S}_i \cdot \boldsymbol{S}_j + Da^3 \sum_{i<j} \left( \frac{\boldsymbol{S}_i \cdot \boldsymbol{S}_j}{|\boldsymbol{r}_{ij}|^3} - \frac{3(\boldsymbol{S}_i \cdot \boldsymbol{r}_{ij})(\boldsymbol{S}_j \cdot \boldsymbol{r}_{ij})}{|\boldsymbol{r}_{ij}|^5} \right) \quad (1)$$

Here $\boldsymbol{S}_i$ represent the Ising spin at each Dy site, $\boldsymbol{r}_{ij}$ are the inter-site distances, $J \approx 1.1$ K is the exchange energy, $D = \mu_0\mu^2/(4\pi a^3)$ the nearest-neighbor dipole interaction energy, and $a$ is the nearest-neighbor distance between moments. From Eqn. 1, only six possible ground-state spin configurations exist on each tetrahedron, all being 2-in/2-out spin arrangements (*23*). Although the dipole interactions in Eqn. 1 could (*24*) generate a first order phase transition to a long-range magnetic ordered state, no signature of such a state has ever been observed to temperatures below $T \approx 50$ mK (*25*). Hence, the monopole kinetics in spin-ice as $T \rightarrow 0$ also remain a focus of concentrated research (*17-21*).

**Supercooling a monopole fluid**

3 By contrast, the excited states governed by Eqn. 1 at higher temperatures $T \gtrsim 1.5$ K, are well understood (*26, 27, 28*) to be mobile magnetic charges (monopoles) of both signs: $+m$ for 1-in:3-out and $-m$ for 3-in:1-out (*SI Appendix,* section I). They exist in a magnetic-charge neutral fluid in which equal numbers of $+m$ and $-m$ are thermally excited across the Dy spin-flip energy barrier $\Delta \approx 4.3$ K. However, below $T \approx 1.5$ K this monopole fluid enters a supercooled state (*29*). Here, the magnetic susceptibility $\chi(\omega, T)$ exhibits a Havriliak-Negami (HN) form (*29*,*30*) characteristic of supercooled glass forming liquids. Further, the susceptibility-derived relaxation time $\tau_\chi(T) = A\exp\left(DT_g/(T - T_g)\right)$ where $D$ is the fragility index, evolves with $T_g \approx 240$ mK $\pm$ 30 mK on a Vogel–Tammann–Fulcher (VTF) trajectory (*29*) characteristic of supercooling (*SI Appendix,* section II). Additionally, Monte Carlo simulations (*31*) predicting magnetization noise with spectral density $S_M(\omega, T) \propto \tau_N(T)/(1 + (\omega\tau_N(T)^b)$ led to the discovery (*32*) of magnetic monopole noise exhibiting a power-law exponent $b(T) \approx 1.5$ approaching $T \approx 1$ K and noise-derived relaxation time $\tau_N(T)$ evolving on an equivalent VTF trajectory *(32,33,34*). Because this is consistent with advanced monopole transport theories based on fractal clusters of monopole trajectories (*19*), heterogeneous monopole dynamics is construed. Altogether, the observed broad distribution of $\chi(\omega, T)$ relaxation times $\tau_\chi(T)$ (*29*), the VTF form evidenced by $\tau_\chi(T)$ (*29*), and the monopole noise power-law $b(T)$ (*32*), imply by analogy with general supercooled glass-forming liquids that monopole dynamical heterogeneity should exist in $Dy_2Ti_2O_7$.

4 Microscopic theories (*17-21*) have long focused on frustrated monopole kinetics approaching the $T \rightarrow 0$ state of spin-ice. Typically, the high-temperature state is viewed as a thermally activated plasma of quasi-free (anti)monopoles (*26, 27, 28*) (state I). Refrigeration from state I is anticipated to yield a supercooled monopole fluid (*29*) (state II) potentially sustaining some form of dynamical heterogeneity (*19,20,21*). The ultra-low temperature state of spin ice (state III) remains empirical *terra incognita*. For example, extended spin-ice models predict growing dynamical heterogeneity resulting in loss of ergodicity near $T/J \approx$ 0.1 when spin-spin correlation time evolves (*17*). Similarly, dumbbell spin-ice models predict that enhancing dynamical heterogeneity near $T \approx 400$ mK in $Dy_2Ti_2O_7$, should cause the fluctuation-dissipation ratio $\omega S_M(\omega,T)/T\chi''(\omega,T)$ to evolve from its ergodic high-temperature limit (*20*). However, the empirical phenomenologies of monopole dynamics in states II and III of spin-ice are largely unexplored. Recent theoretic advances actualize these concepts by predicting a new form of heterogeneous monopole dynamics based on the existence of two spin-dynamical time-scales, constraining the trajectories of monopoles to nanoscale clusters (*19*), and this hypothesis is supported indirectly by recent experiments (*32,33,34).*

**5** If its microscopic kinetics were not frustrated, dipolar spin ice is predicted to undergo a first order transition to a magnetically long-range ordered phase (*24*). Yet this ordered state has never been observed in $Dy_2Ti_2O_7$ or $Ho_2Ti_2O_7$ apparently due to frustration of the necessary global spin reconfigurations by a network of Dirac-strings threading the material, each string being due to passage of a magnetic (anti)monopole. The consequent spin ice 'clusters' are conceived as local regions of 2-in:2-out spin configurations that have net coarse-grained magnetization. Within each, the motion of (anti)monopoles flips one spin per tetrahedron all along its Dirac-string trajectory, resulting in a different configuration of 2-in:2-out spins due to the passage of each monopole. However, once all internal monopole motion has halted, the net magnetic moment of such a cluster is unchanged. By definition, such a reconfiguration of 2-in:2-out spin arrangements within a spin ice cluster corresponds to a burst of monopole motion. Spin ice theories describe that process colloquially as e.g. "large-scale annealing events where many monopoles exchange position in bursts and the effective potential landscape changes akin to an avalanche" (*21*). Fundamentally, monopole current bursts and spin ice cluster reconfigurations are synonyms for the same microscopic concept: one cannot have the reorganization of the underlying 2-in:2out spin patterns (the spin ice cluster configuration) without the simultaneous movement of the monopoles (the monopole current).

### Coterminous monopole noise and magnetic susceptibility

6 To explore such concepts in $Dy_2Ti_2O_7$, we use SQUID-based flux-noise spectrometry with magnetic field sensitivity $\delta B = \mu_0 \delta M \leq 10^{-14}$ T/$\sqrt{\mathrm{Hz}}$, where $\mu_0$ is the permeability of vacuum. Using the apparatus shown schematically in Fig. 1A. Here $L_p$ is the inductance of both the sample pickup coil and of a counter wound compensation coil, $L_i$ is a SQUID-input coil inductance, and $\mathcal{M}_i$ is a mutual inductance to SQUID. Our spectrometer is operated on a cryogen-free dilution refrigerator in the range 15 mK $\lesssim T \lesssim$ 2500 mK. The time-sequence of the magnetic flux generated by the sample, $\Phi_p(t,T)$, is measured at each temperature T with microsecond precision, using a persistent superconducting circuit that transforms it into the flux $\Phi(t,T)$ at the SQUID input coil

$$\Phi(t,T) = (\mathcal{M}_i/(2L_p + L_i))\Phi_p(t,T) = \Phi_p(t,T)/\beta \qquad (2)$$

where $\beta$ is a precisely and independently known calibration constant of the instrument (*SI Appendix,* section III). Although the procedures for analyzing magnetic monopole noise from time sequences of magnetic flux $\Phi_p(t,T)$ are well established (*32,33,34*), we review them here for didactic purposes. As illustrated in Fig. 1B, the time dependence of sample magnetization along the symmetry axis of the superconductive pickup coil is

$$M(t,T) \equiv \Phi_p(t,T)/A\mu_0 \qquad (3)$$

where A is the sample cross-sectional area. The time dependence of the component of magnetic monopole current $J(t,T)$ (Fig. 1C) along the same axis of the pickup coil is

$$J(t,T) \equiv \dot{\Phi}_p(t,T)/\mu_0 \qquad (4)$$

The total magnetic energy $\varepsilon(t,T)$ (Fig. 1D) due to the monopole currents which is stored in the pickup coil at any instant is

$$\varepsilon(t,T) \equiv \Phi_p^2(t,T)/2L_p \qquad (5)$$

The power spectral density of the magnetization fluctuations from Eqn. 3 is

$$S_M(\omega,T) \equiv \lim_{P\to\infty} \frac{1}{P}\left|\int_{-\frac{P}{2}}^{\frac{P}{2}} M(t,T)e^{\,i\omega t}dt\right|^2 \tag{6}$$

where $P$ is the time period over which the time sequence of $M(t,T)$ is recorded. The noise contribution of the superconductive circuitry and SQUID as measured separately with no sample, are first subtracted to yield the sample $S_M(\omega,T)$. Finally, the magnetic susceptibility $\chi(\omega,T)$ is the magnetic response of the sample to an applied field $H$ at angular frequency $\omega$. The in-phase and out-of-phase response of magnetic flux from the sample only, $\Phi_p^X(\omega,T)$ ; $\Phi_p^Y(\omega,T)$, as measured using a calibrated lock-in amplifier to analyze $\Phi_p(t,T)$, yields the in-phase and out of-phase susceptibility as

$$\chi(\omega,T) \equiv \chi'(\omega,T) + i\chi''(\omega,T) = \frac{1}{NA}\left[\frac{\Phi_p^X(\omega,T)}{\mu_0 H} + i\,\frac{\Phi_p^Y(\omega,T)}{\mu_0 H}\right] \tag{7}$$

where N is the number of turns in the superconductive pickup coil. This magnetic susceptibility $\chi(\omega,T)$ is measured coterminously with the magnetization noise $S_M(\omega,T)$, using identical samples / detectors, over the temperature range 250 mK $< T <$ 2500 mK. As temperature falls, the monopoles/antimonopole pairs become dilute (*SI Appendix,* section I) with approximate density $n \approx n_0 \exp(-\Delta/\mathrm{kT})$ where $n_0 = 1/a_d \approx 10^{28}\mathrm{m}^{-3}$ and $\Delta \approx$ 4.3 K. Thus at $T \approx 250$ mK, $n \approx 10^{20}\ \mathrm{m}^{-3}$ so that for sample volume $\upsilon \approx 10^{-8}\mathrm{m}^3$, $n \approx 10^{12}$ pairs remain. This number is well above out detection threshold. All the above analysis procedures are exemplified in Fig. 1C, D, based on a typical time sequence of flux noise $\Phi_p(t,T)$ in Fig. 1B, and all subsequent studies reported below are based on such data.

7 For an ergodic monopole fluid, the fluctuation-dissipation theorem (FDT) linking $S_M(\omega,T)$ to the imaginary magnetic susceptibility $\chi''(\omega,T)$ would predict (*20*)

$$S_M(\omega,T) = 2k_\mathrm{B}T\,\chi''(\omega,T)/\omega\pi\upsilon\mu_0 \tag{8}$$

where $\upsilon$ is the sample volume, $k_\mathrm{B}$ is Boltzmann's constant and we use SI units throughout. Conversely, a violation of FDT would imply a loss of ergodicity in the monopole fluid. For our $Dy_2Ti_2O_7$ samples, a typical set of coterminously measured $\chi'(\omega,T)$, $\chi''(\omega,T)$ and $S_M(\omega,T)$ are plotted in Fig. 1E (*SI Appendix,* section IV). Here, because of the wide distribution of microscopic relaxation times (*29*), even when $\tau_{\chi}(T)$ increases to the glass transition at $T_g \approx$ 250 mK (*29*), high frequency monopole dynamics must still be present at a subset of sites. Hence, to explore the evolution of Eqn. 8 to lowest temperatures, we plot in Fig. 1F the measured $S_M(\omega, T)$ versus independently measured $2k_{\mathrm{B}}T\,\chi''(\omega, T)/\omega\pi\upsilon\mu_0$ at frequencies where dynamics is manifestly occurring in the monopole noise. Evidently, the fluctuation-dissipation theorem for the magnetic monopole fluid holds for $T \gtrsim 500$ mK. However, because of the observed collapse of $X(\omega, T) \equiv S_M(\omega, T)\omega\pi\upsilon\mu_0\, /\, 2k_{\mathrm{B}}T\,\chi''(\omega, T)$ from $X = 1$ starting below $T \lesssim 500\,\mathrm{mK}$, the monopole fluid here slowly exits the ergodic regime. Eventually FDT is strongly violated with complete loss of monopole ergodicity $T \lesssim 250$ mK as shown Fig. 1F (*SI Appendix,* section V).

### Discovery of monopole dynamical heterogeneity

8 A key signature of the dynamical component of monopole dynamical heterogeneity would be random and intense monopole current bursts (*19, 21*). Hence, we next measure the time-sequences of flux threading the sample at its pickup coil, $\Phi_p(t, T)$, typically for a continuous period of $P = 1000$ seconds. If each monopole exhibits a magnetic charge $m$ and total magnetic flux $\Phi_m = m\mu_o$ (*26*) and because the magnetic flux through any superconductive closed-loop circuit is quantized, when a magnetic monopole passes through such a loop it generates a supercurrent exactly counterbalancing $\Phi_m$. This is detectable by a SQUID as a flux generated elsewhere in the circuit. Under these circumstances, the time dependence of net monopole current through the pickup coil is $J(t, T) \equiv \dot{\Phi}_p(t, T)/\mu_0$ (*SI Appendix,* section VI). For such measurements of $J(t, T)$ derived from the unprocessed $\Phi_p(t, T)$ data, we use an 80 μs box-car average, with an exemplary time sequence of $J(t, T)$ as derived from $\Phi_p(t, T)$ in Fig. 1B, being shown in Fig. 1C. Typical measured time sequences of monopole current magnitudes $|J(t, T)|$ derived using Eqn. 4, over a wide range of temperatures traversing from the homogeneous monopole fluid regime into the supercooled regime, are shown in Fig 2A. The measured probability distribution, $r_{|J|}$, of the magnitudes of monopole currents $|J(t,T)|$ during $P = 1000$ second periods, is given by the number $\eta(|J|)$of monopole currents per unit time having magnitude $|J|$: $r_{|J|} = \eta(|J|)/P$. A typical set of distributions of $r_{|J|}$ is then shown in Fig. 2B. Here the monopole current magnitudes range in intensity over almost five orders of magnitude with maximum intensity occurring near $T = 1$ K. The evolution of $r_{|J|}$, is presented in Fig. 2C versus T. The time-averaged intensity of these same monopole currents $\overline{|J|}(T)$ is shown in Fig. 2D.

9 As to the energetics of these phenomena: Fig. 1B provides a typical example of magnetization fluctuations in terms of $\Phi_p(t,T)$ and the magnetic energy $\varepsilon$ associated with each global monopole configuration can then be determined accurately from the unprocessed $\Phi_p(t,T)$ signal. From elementary superconductive circuit analysis, $\varepsilon(t,T) \equiv \Phi_p^2(t,T)/2L_p$ (*SI Appendix,* section VI). Typical examples of measured values of $\Phi_p^2(t,T)$ during 1000 second periods are shown in Fig. 3A for a representative set of temperatures. The measured probability distribution $r_\varepsilon(T)$, of global monopole dynamic events with energy $\varepsilon$ per unit time, is given by the number $m(\varepsilon)$ of times a given energy $\varepsilon$ occurs in the continuous energy signal: $r_\varepsilon = m(\varepsilon)/P$. Typical histograms of $r_\varepsilon(T)$ are presented in Fig. 3B versus *T.* Strikingly, while the energetics $\varepsilon(t)$ are gaussian and narrow in distribution for $T \gtrsim 1500$ mK, at lower temperatures a sharp bifurcation occurs into a bi-modal distribution containing less frequent highly energetic events, each exemplifying a monopole-current burst. Eventually below $T \lesssim 250$ mK these phenomena disappear, and a low energy gaussian distribution reappears. This complete phenomenology is represented by all measured $r_\varepsilon(T)$ data shown as a color-coded 2D histogram in Fig. 3C. Here, the dashed curve $\bar{\varepsilon}_M(T)$ indicates the average energy of conventional monopole generation-recombination noise (*32,33,34*) while the dotted curve $\bar{\varepsilon}_B(T)$ plots the average energy of monopole current bursts ascribed to dynamical heterogeneity. The measured relative energy intensities of monopole current bursts $\bar{\varepsilon}_B(T)$ and of $\bar{\varepsilon}_M(T)$ are shown in Fig. 3D. Clearly, there are two populations of monopole currents: those related to conventional monopole noise (*32,33,34*), plus more intense current bursts existing over extended time periods and producing large excursions in $\Phi_p(t,T)$ (e.g. arrows in Fig. 2A). A strong maximum in monopole current burst intensity occurs entering the supercooled regime, followed by a rapid collapse below $T \lesssim 500$ mK (Fig. 2D). The ostensible cause of this bimodality in monopole currents is that elementary monopole generation recombination processes (*32,33,34*) fall into one class of monopole dynamics, whereas the monopole current bursts causing dynamical reconfiguration of larger spin ice clusters represents a far more intense second class (*19,21*).

**10** The relaxation time of magnetization fluctuations $\tau_N(T)$ in the supercooled regime of $Dy_2Ti_2O_7$ is determined via its power spectral density $S_M(\omega,T)$, by fitting to $S_M(\omega,T) \propto \tau_N(T)/(1+(\omega\tau_N(T)^b)$ over almost four orders of magnitude in $\tau_N(T)$ (*SI Appendix,* section VII). As had previously been established from susceptibility studies (*29*), here we find that $\tau_N$ (T) for $Dy_2Ti_2O_7$ is governed demonstrably by the Vogel-Tammann-Fulcher (VTF) equation $\tau_N(T) = A\exp(DT_g/T - T_g)$, with fragility $D \approx 14$ and a well-defined glass temperature $T_g \cong 240$ mK (*SI Appendix*, section VII), while sustaining ergodicity at all higher temperatures. Equivalent relaxation time characteristics evolving according to the VTF equation are widely observed in molecular glass forming liquids (*2-8*).

**11** Finally, while the dynamical nature of the monopole current bursts is self-evident (Figs. 1-3) their heterogeneity requires quantification. In the general theory of dynamical heterogeneity in supercooled liquids (*2-8*), slower dynamics continuously transform to faster dynamics and vice versa at ever changing nanoscale regions with correlation length $\xi(T)$, a spatial scale that increases rapidly towards the glass transition. The empirical challenge for monopole fluid studies is then to characterize such coterminous phenomena in terms of their lengthening lifetimes (which are well established for $Dy_2Ti_2O_7$ (*28*, *29*,*32*), and of their increasing length scales $\xi(T)$ which are unknown. In principle, the latter may be determined by using the four-point susceptibility $\chi_4(\tau,T)$ (*35*,*36*,*37*), a measure of the fluctuations in the two-point correlation function. Theories of glass formation report that $\chi_4(\tau,T)$ typically exhibits a strong maximum in relaxation time $\tau$ dependence, whose height is proportional to the volume containing the correlated motion of molecules (*38-41*). Direct experimental measurements of $\chi_4(\tau,T)$ have focused on non-thermodynamic colloidal and granular materials (*42,43,44*) where heterogeneity is accessible experimentally by imaging. But direct measurements of $\chi_4(\tau,T)$ are very challenging for physical systems containing an ensemble of nanoscale particles in thermodynamic equilibrium, e.g. supercooled glass forming molecular liquids.

**12** To circumvent this limitation, another approach has been developed (*13,37,45,46*). Given a time-series of measurements $A(t,T)$, where $A$ is a property of a system in thermal equilibrium at temperature $T$, the standard two-point correlation function is $C_A(t,\tau,T) = A(t,T)A(t+\tau,T)$ while the consequent autocorrelation function is $F_A(\tau,T) \equiv \langle A(t,T)A(t+\tau,T)\rangle_t$ (*SI Appendix,* section VIII). The dynamic susceptibility $\chi_4(\tau,T)$ can then be estimated (*13,37,45,46*) from the response function $\chi_T(\tau,T)$ to temperature variations which is defined as

$$\chi_T(\tau,T) = \partial F_A(\tau,T)/\partial T \quad (11)$$

When the fluctuation-dissipation relation is valid, it has then been established (*45*) that $k_B T^2 \chi_T(\tau,T) = N\langle \delta C_A(t,\tau,T)\delta H(t,0,T)\rangle_t$ where $\delta C_A(t,\tau,T) = C_A(t,\tau,T) - \langle C_A(\tau,T)\rangle_t$ is the fluctuation of $C_A(t,\tau,T)$ about its mean value $\langle C_A(\tau,T)\rangle_t$, while $\delta H(t,0,T)$ is the fluctuating enthalpy per particle, and $N$ is the total number of particles. Experimental and numerical studies (*37,45*) have shown that from these concepts it follows

$$\chi_4(\tau,T) \approx k_B T^2 c_p(T)^{-1}[\,\chi_T(\tau,T)]^2 \quad (12)$$

where $c_p(T)$ is the specific heat capacity of the particles undergoing vitrification. In the context of $Dy_2Ti_2O_7$ magnetic monopole fluids where $\Phi_p(t,T)$ is the thermodynamic property fluctuating in time, the correlation function and autocorrelation functions have been established previously (*32*): the relevant two-point correlation function is $C_\Phi(t,\tau,T) = \Phi_p(t,T)\Phi_p(t+\tau,T)$ while its autocorrelation function is $F_\Phi(\tau,T) \equiv \langle\Phi_p(t,T)\Phi_p(t+\tau,T)\rangle_t$ . The four-point dynamic susceptibility $\chi_4(\tau,T)$ of the monopole fluid can then be estimated using Eqn. 12. To explore this concept, the normalized autocorrelation function $F_\Phi(\tau,T)$ is calculated from the unprocessed $\Phi_p(t,T)$ data (in a $P = 1000$ second interval) as

$$F_\Phi(\tau,T) = N_F(T)\frac{1}{P-\tau}\sum_{t=0}^{P-\tau}\Phi_p(t,T)\Phi_p(t+\tau,T) \tag{13}$$

where $N_F(T) = 1/(\sum_{t=0}^{P}\Phi_p(t,T)\Phi_p(t,T))$ is the normalization constant ensuring $F(\tau = 0,T) = 1$. A 2D plot of $F_\Phi(\tau,T)$ as determined using Eqn. 13 is generated and then further interpolated in 10 mK temperature steps (*SI Appendix*, section VIII). The response function $\chi_T(\tau,T) = \partial F_\Phi(\tau,T)/\partial T$ is then calculated directly from that $F_\Phi(\tau,T)$ . Finally, the four-point dynamic susceptibility $\chi_4(\tau,T)$ of the $Dy_2Ti_2O_7$ monopole fluid is estimated from $\chi_T(\tau,T)$ using Eqn. 12 with specific heat data $c_p(T)$ as previously determined (*47*) (*SI Appendix,* section IX). Fig. 4A presents the resulting $\chi_4(\tau,T)$ determined within the supercooled monopole fluid regime. It immediately reveals the increasing intensity in the evolution of the maxima in $\chi_4(\tau,T)$ with falling temperature. These characteristics are patently consistent with long-established theory for $\chi_4(\tau,T)$ in glass-forming molecular liquids (*35,36,37*) wherein, if dynamical heterogeneity is spatially compact, evolution of its length scale is then given by $\xi(\mathrm{T}) \propto \sqrt[3]{MAX(\chi_4(\tau,T))}$ (*38-41,48,49*). In consequence, Fig. 4B represents the measured temperature evolution of the maxima of $\chi_4(\tau,T)$ and hence the evolving length scales of dynamical heterogeneity in a supercooled monopole fluid, with $\bar{\xi}(T) \equiv \xi(\mathrm{T})/\xi(T = 1.5\text{ K})$ increasing by almost a factor of 8 across the supercooled regime. Furthermore, the time-evolution of the maxima in $\chi_4(\tau,T)$ reveals a dramatic slowing of the dynamical heterogeneity. The time over which the dynamics retain maximum spatial correlation is the characteristic dynamical heterogeneity time $\tau_4$ (*39*). In theory, as the correlation length evolves approaching a glass transition, the relaxation time too must evolve as ever-larger regions of the material must rearrange cooperatively making such rearrangements exponentially rarer. For $Dy_2Ti_2O_7$ the relaxation times $\tau_4(T)$ are determined by measuring the times at which $\chi_4(\tau, T)$ is maximum for each temperature. In Fig. 4C, these times are compared to the relaxation times $\tau_N(T)$ derived from fitting the monopole noise spectrum to $S_M(\omega, T) \propto \tau_N(T)/(1 + (\omega\tau_N(T)^b)$, which themselves are known to be largely consistent with the relaxation times $\tau_\chi(T)$ from susceptibility measurements (*29*) (*SI Appendix*, section VII). All three independently determined relaxation times: $\tau_4(T), \tau_N(T)$ and $\tau_\chi(T)$, are in good agreement for $0.5\ \mathrm{K} < T < 1.5\ \mathrm{K}$, thus revealing that the well-known VTF evolution of relaxation times $\tau_\chi(T) \approx \tau_N(T) = A\exp\left(DT_g/(T - T_g)\right)$ in $Dy_2Ti_2O_7$, is due to dynamical heterogeneity.

**13** We amalgamate all the above results on the emerging phenomenology of dynamical heterogeneity in $Dy_2Ti_2O_7$ spin-ice in Fig. 4D. Below $T \approx 1500$ mK, intense monopole current bursts emerge indicating large scale reorganizations of spin ice configurations. Their maximum magnitude relative to the generation/recombination monopole noise $\mathcal{R} = \max(\varepsilon_B)/\overline{\varepsilon_M}$ grows rapidly, reaching maximum near $T \approx 750$ mK and eventually disappears near $T_g \approx 250$ mK (Fig. 4D (i)). This novel bimodality is due to elementary monopole generation recombination processes (*32,33,34*) falling into one class of monopole dynamics, with the monopole current bursts from dynamical reconfiguration of spin ice clusters being an intense second class (*19,21*). Traversing this supercooled regime, a direct measure of monopole ergodicity $X(\omega, T)$ diminishes cumulatively, reaching a minimum at $T \lesssim T_g$ (Fig. 4D (ii)). Across the same regime the power law of magnetization noise collapses from the expected (19) value $b$=1.5 for quasi-free monopoles, toward $b$=1 approaching $T_g$ (Fig. 4D (iii)). The relative dynamical heterogeneity length scale $\bar{\xi}(T)$ increases significantly across the supercooled regime so that the volume of dynamically heterogeneous regions increases by a factor of approximately 500 as $T_g$ is approached (Fig. 4D (iv)). Overall this provides a far clearer and more comprehensive understanding of the evolution of spin ice dynamics from the thermally activated monopole plasma (state I), through the supercooled monopole fluid (state II) reaching the glass transition at $T_g \approx 250$ mK, and into that ultra-low temperature state (state III).

**Exegesis**

**14** More generally, the striking correspondence between the phenomenology of dynamical heterogeneity of supercooled monopole fluids (Figs. 2-4) and that in supercooled glass forming liquids (*2-8*) emphasizes a potential universality among these ostensibly very distinct microscopic phenomena. In this regard, classical spin ice may provide a physically realized version of kinetically constrained models (KCM) for facilitated, heterogeneous relaxation during vitrification (*50*). But, unlike conventional KCMs where dynamical rules are postulates and their consequences then explored, the microscopic kinetic constraints of spin ice are directly identifiable. This is because, starting from a simple, clear and quantitative Hamiltonian in Eqn. 1, the kinetic constraints on the spatiotemporal relaxation paths of spins emerge naturally from monopole motion through the dynamically evolving Dirac string network. The presence of such kinetic constraints in spin ice manifestly induces a slowing of spin relaxation, the subsequent appearance of dynamical heterogeneity as reported here, followed by an eventual glass state at $T_g \approx 250\,\text{mK}$. Thus, in frustrated molecular glass-forming systems, the relaxation is plausibly controlled by rare, spatially-correlated geometrical reconfigurations that become increasingly improbable with falling temperature while, in frustrated monopole glass-forming systems, the relaxation is controlled by rare, spatially-correlated monopole reconfigurations of the 2-in:2-out spin arrangements, that become increasingly improbable with falling temperature. If correct, this represents a remarkable new physics opportunity because spin ice provides a uniquely transparent realization of facilitated vitrification dynamics in which the kinetic constraints are microscopically well understood (*17-22*) and their spatiotemporal dynamical consequences are now directly accessible experimentally (Figs. 1-4).

**FIGURES**

Figure 1

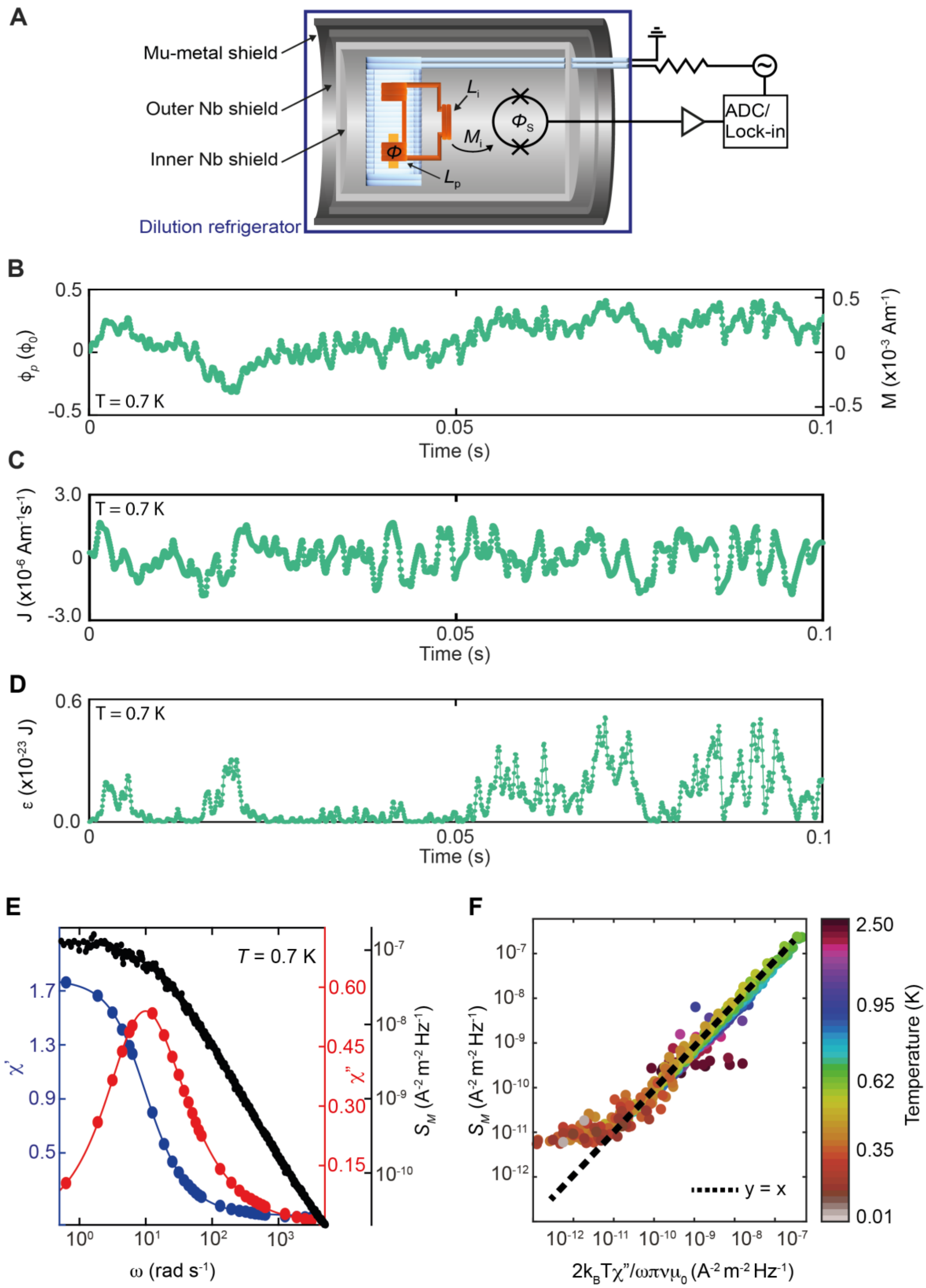

A
Mu-metal shield
Outer Nb shield
Inner Nb shield
Dilution refrigerator
L_i
M_i
Φ_S
Φ
L_p
ADC/
Lock-in
B
φ_p (φ_0)
M (x10^-3 Am^-1)
T = 0.7 K
Time (s)
C
J (x10^-6 Am^-1s^-1)
T = 0.7 K
Time (s)
D
ε (x10^-23 J)
T = 0.7 K
Time (s)
E
T = 0.7 K
χ'
χ"
S_M (A^-2 m^-2 Hz^-1)
ω (rad s^-1)
F
S_M (A^-2 m^-2 Hz^-1)
2k_BTχ"/ωπνμ_0 (A^-2 m^-2 Hz^-1)
y = x
Temperature (K)
2.50
0.95
0.62
0.35
0.01

**Figure 1 Magnetic monopole noise spectrometry.**

A. Schematic of the experimental apparatus we use for detection of dynamical heterogeneity due to magnetic monopole current bursts in the supercooled monopole fluid of $Dy_2Ti_2O_7$.

*B.* Typical example of the magnetic flux $\Phi_p(t,T)$ generated spontaneously by $Dy_2Ti_2O_7$ samples. The time dependence of sample magnetization along the symmetry axis of the superconductive pickup coil is the $M(t,T) = \Phi_p(t,T)/A\mu_0$ from Eqn. 3, as shown on RHS.

C. Typical time dependence of magnetic monopole current $J(t,T) \equiv \dot{\Phi}_p(t,T)/\mu_0$ along the axis of the pickup coil from Eqn. 4.

*D.* Typical total magnetic energy $\varepsilon(t,T) \equiv \Phi_p^2(t,T)/2L_p$ due to the monopole currents within the sample as detected by the pickup coil at any instant *t,* from Eqn. 5.

*E.* The power spectral density of the $Dy_2Ti_2O_7$ sample magnetization fluctuations $S_M(\omega,T) \equiv \lim_{P\to\infty} \frac{1}{P}\left|\int_{-\frac{P}{2}}^{\frac{P}{2}} M(t,T)e^{\,i\omega t}dt\right|^2$, where P is the time period over which the time sequence of $M(t,T)$ is recorded as in B. Typical $Dy_2Ti_2O_7$ magnetic susceptibility $\chi'(\omega,T)$, $\chi''(\omega,T)$ and magnetization noise spectrum $S_M(\omega,T)$ measured coterminously at $T =$ 700 mK.

F. Temperature dependence of coterminous $S_M(\omega,T)$ and $\chi''(\omega,T)2kT/\omega\pi\upsilon\mu_0$. Evidently, monopole ergodicity parameterized by $X(\omega,T) \equiv S_M(\omega,T)/\{\chi''(\omega,T)2kT/\omega\pi\upsilon\mu_0\}$ diminishes slowly beginning near $T \approx 500$ mK, to be lost manifestly by $T \lesssim 250$ mK.

Figure 2

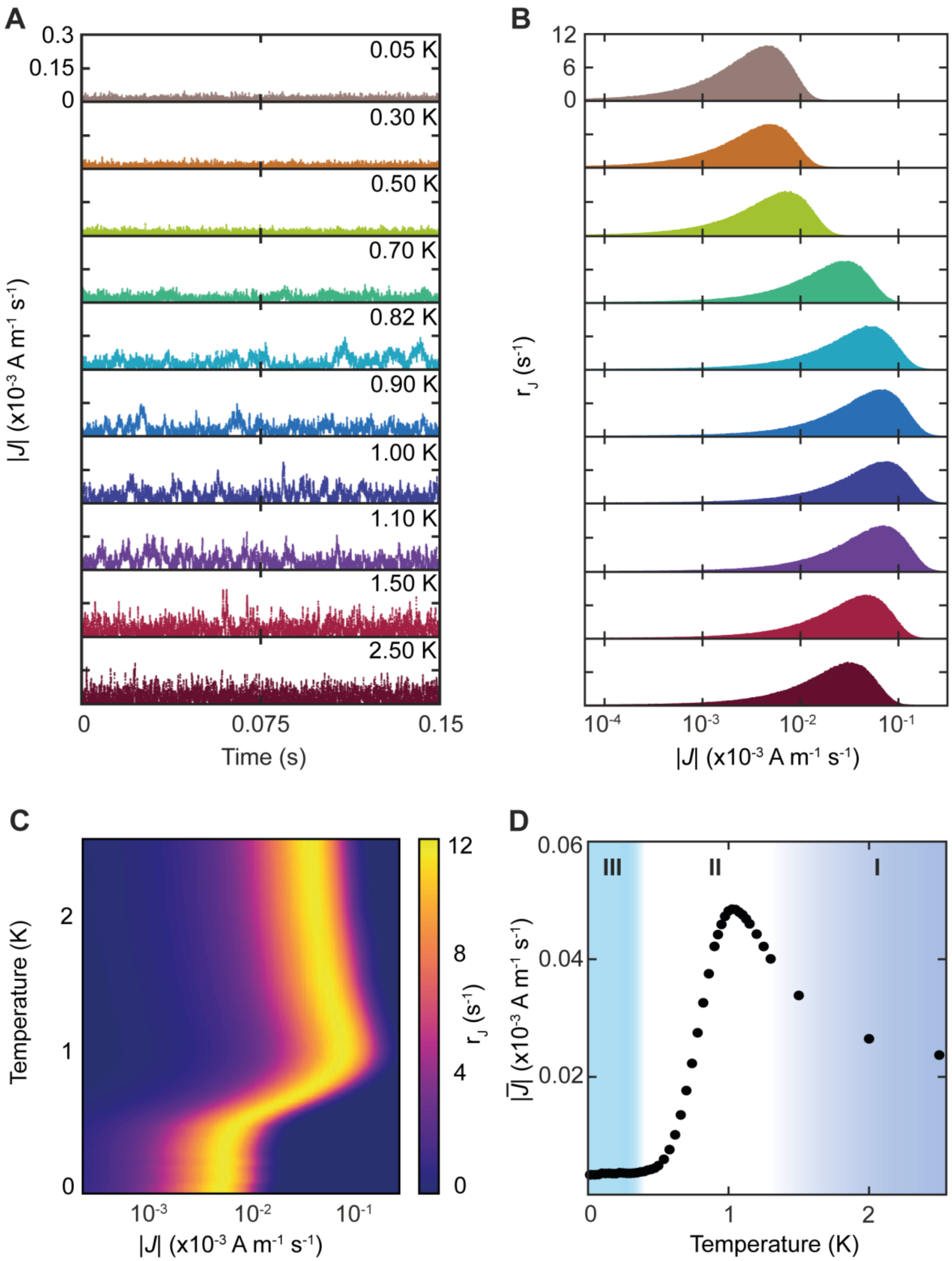

A
B
C
D
0.05 K
0.30 K
0.50 K
0.70 K
0.82 K
0.90 K
1.00 K
1.10 K
1.50 K
2.50 K
0.3
0.15
0
|J| (x10^-3 A m^-1 s^-1)
0
0.075
0.15
Time (s)
12
6
0
r_J (s^-1)
10^-4
10^-3
10^-2
10^-1
|J| (x10^-3 A m^-1 s^-1)
Temperature (K)
2
1
0
12
8
4
0
r_J (s^-1)
III
II
I
0.06
0.04
0.02
|J̄| (x10^-3 A m^-1 s^-1)
Temperature (K)

**Figure 2 Monopole current bursts in the supercooled state.**

A. Typical measured time sequences of monopole current magnitudes $|J(t,T)|$ from Eqn. 4, over a wide range of temperatures spanning the homogeneous monopole fluid regime I, into the supercooled regime II, and finally the $T \rightarrow 0$ regime III.

B. Typical measured probability distribution of the monopole current burst magnitudes $|\ J(t,T)| = |\dot{\Phi}(t,T)/\mu_0|$ derived directly from the unprocessed and non-normalized $\Phi(t,T)$ data, e.g., in A. The measured monopole currents span an intensity range of approximately five orders of magnitude, with maximum intensity individual events occurring at $T \approx 900$ mK. These data are highly typical of multiple $Dy_2Ti_2O_7$ samples studied.

C. Typical per unit time rate $r_{|J|}$ of monopole currents having magnitude $|J|$, measured versus temperature $T$. The rate of occurrence $r_{|J|}$ of a monopole current with magnitude $|J|$ is defined as the number $\eta(|J|)$ observed in given time interval P: $r_{|J|} \equiv \eta(|J|)/P$.

D. Average measured intensity of monopole current bursts $\overline{|J|}$ versus temperature. Clearly, approaching the supercooled regime below $T \approx 1500$ mK they intensify dramatically, only to fall precipitously reaching a plateau $T \lesssim 250$ mK.

Figure 3

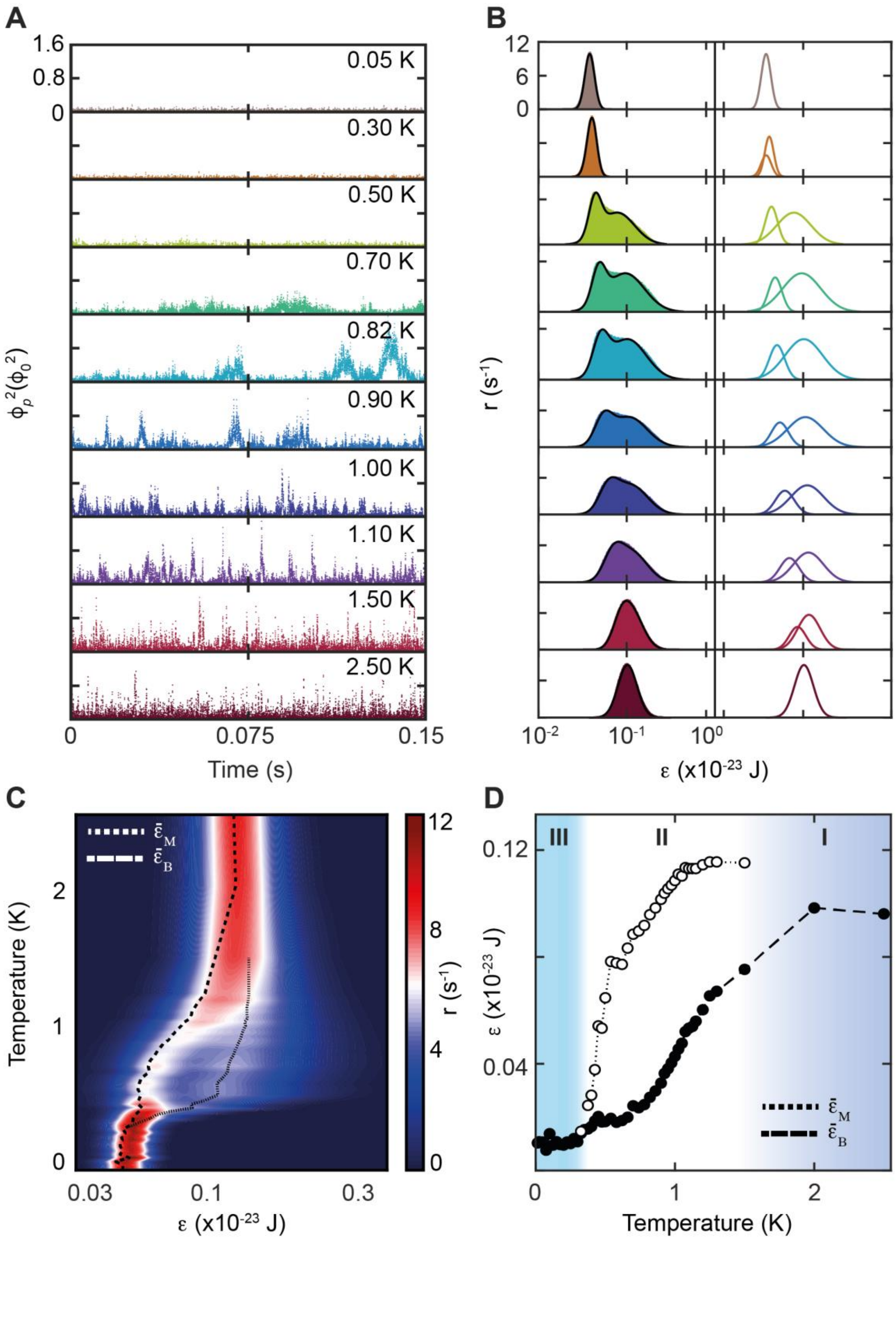
A
1.6
0.8
0
0.05 K
0.30 K
0.50 K
0.70 K
0.82 K
0.90 K
1.00 K
1.10 K
1.50 K
2.50 K
$\phi_p^2(\phi_0^2)$
0
0.075
0.15
Time (s)
B
12
6
0
r (s$^{-1}$)
$10^{-2}$
$10^{-1}$
$10^{0}$
ε (x10$^{-23}$ J)
C
$\bar{\varepsilon}_M$
$\bar{\varepsilon}_B$
Temperature (K)
2
1
0
0.03
0.1
0.3
ε (x10$^{-23}$ J)
12
8
4
0
r (s$^{-1}$)
D
III
II
I
0.12
0.04
ε (x10$^{-23}$ J)
$\bar{\varepsilon}_M$
$\bar{\varepsilon}_B$
0
1
2
Temperature (K)

**Figure 3 Monopole noise bifurcation due to dynamical heterogeneity.**

A. Typical examples of the $\Phi_p^2(t,T)$ from directly measured time dependence of spontaneous magnetic flux $\Phi_p(t,T)$. This is shown, for example, at temperatures 50 mK, 250 mK, 500 mK, 700 mK, 900 mK, 1500 mK, and 2500 mK.

B. Typical histograms of the measure rate of flux states $r_\varepsilon(T)$ versus ε. We define the rate of occurrence $r_\varepsilon(T)$of any state with energy $\varepsilon$ as the number $m(\varepsilon)$ observed in given time interval $P$: $r_\varepsilon \equiv m(\varepsilon)/P$. Conventional monopole generation-recombination noise with a simple Gaussian distribution persists until $T \approx 1500$ mK. More intense monopole current bursts with far higher energy appear below this temperature resulting in a bimodal distribution of probabilities as shown via histograms at left, and by the fit curves to each histogram shown at right. Eventually below $T \lesssim 250$ mK the bimodal distribution of monopole current burst energies disappears.

C. Monopole noise bifurcation effect in Fig. 3B is presented as a color-coded 2D histogram containing $r_\varepsilon$ versus $T$. Dashed curve $\bar{\varepsilon}_M(T)$ indicates the average energy of conventional monopole noise, while the while the dotted curve $\bar{\varepsilon}_B(T)$ plots the average energy of monopole current bursts ascribed to dynamical heterogeneity.

D. Relative intensities of average energy of monopole current bursts $\bar{\varepsilon}_B(T)$ and of conventional monopole noise $\bar{\varepsilon}_M(T)$.

Figure 4

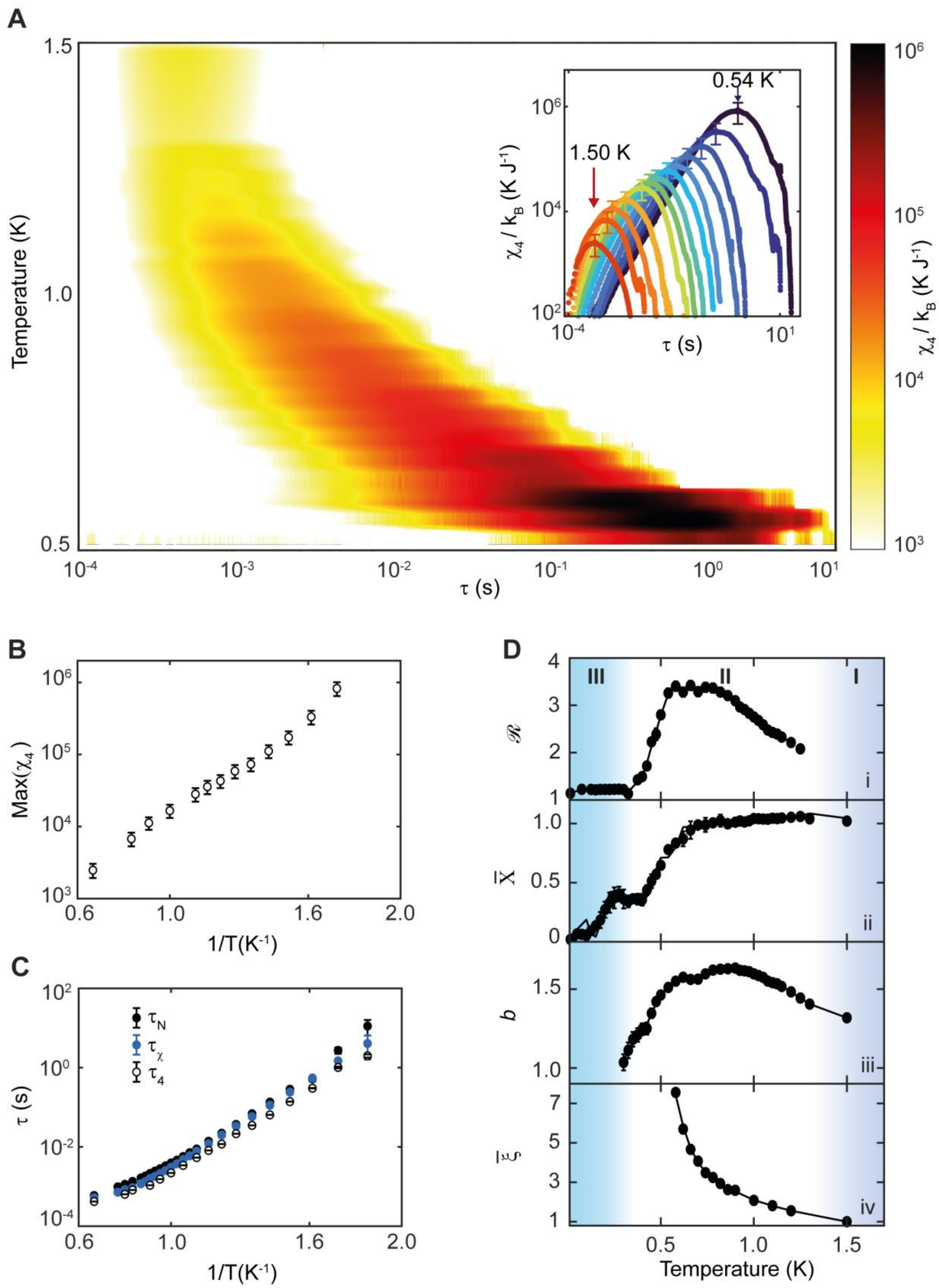

A
Temperature (K)
1.5
1.0
0.5
10^-4
10^-3
10^-2
10^-1
10^0
10^1
τ (s)
10^6
10^5
10^4
10^3
χ4 / kB (K J^-1)
0.54 K
1.50 K
B
Max(χ4)
1/T(K^-1)
0.6
1.0
1.6
2.0
C
τN
τχ
τ4
τ (s)
1/T(K^-1)
D
III
II
I
ℛ
i
ii
iii
iv
b
ξ
Temperature (K)

**Figure 4 Measured $\chi_4(T,\tau)$ and $\bar{\xi}(T)$ of monopole dynamical heterogeneity**

A. Measured dynamical susceptibility $\chi_4(\tau,T)$ of the supercooled monopole fluid of $Dy_2Ti_2O_7$. Inset: $\chi_4(\tau,T)$ shown on a log-log scale at a representative set of temperatures.

B. Evolution of $MAX(\chi_4(\tau,T))$ with temperature shows the striking growth in relative correlation length of dynamical heterogeneity in $Dy_2Ti_2O_7$.

C. Evolution of the relaxation times with inverse temperature 1/T. The filled black circles represent $\tau_N(T)$ extracted by fitting the noise spectra $S_M$. The blue filled circles represent $\tau_\chi(T)$ which is extracted as the peak of the fit to the imaginary magnetic susceptibility. The empty circles represent $\tau_4(T)$ extracted from the time at which dynamical susceptibility $\chi_4(T)$ achieves its maximum.

D. (i) Measured ratio of maximal monopole current bursts relative to the conventional magnetic monopole noise $\mathcal{R} \equiv max\ (\varepsilon_B)/\ \overline{\varepsilon_M}$; (ii) Measured monopole fluid ergodicity $X(\omega,T) = 2k_BT\chi''(\omega,T)/\omega\pi\upsilon\mu_0S_M(\omega,T)$ ; (iii) Measured frequency-dependent power law $b(T)$ of magnetization noise; (iv) Measured evolution of relative correlation length $\bar{\xi}(T)$ of dynamical heterogeneity. Evidently, all four characteristics of magnetic monopole dynamics span the same three ranges of temperature: thermally activated quasi-free monopole fluid (I) indicated in darker blue; the supercooled regime encompassing the monopole dynamical heterogeneity phenomenology (II) in white; and the yet unexplored regime (III) in light blue.

**Acknowledgements**: We acknowledge and thank J. Hallén, C. Castelnovo, S. Giblin, R. Dusad, S.A. Kivelson, Z. Nussinov, O.H. Selby-Davis and S. Sondhi and for key discussions and guidance.

J.C.D., C.D., J. M., J.W. and J.C.S.D. acknowledge support from Research Ireland under Award SFI 17/RP/5445.

C.C acknowledges support from Research Ireland under Award GOIPG/2023/4014.

J.C.S.D. and F.J. thank the Max Planck Institute for Chemical Physics of Solids for support.

S.J.B. acknowledges support from UK Research and Innovation (UKRI under the UK government's Horizon Europe funding guarantee (Grant No. EP/X025861/1).

J.C.S.D. acknowledges support from the Moore Foundation's EPiQS Initiative through Grant GBMF9457.

C.-C.H. and J.C.S.D. acknowledge support from the European Research Council (ERC) under Award DLV-788932.

H.T. and J.C.S.D. acknowledge support from the UK Royal Society under Award R64897.

**Author Contributions:** JCSD and JW conceived the project. GL and SS synthesized and characterized the samples; JCD, FJ, CC, C-CH, HT developed relevant monopole noise spectroscopy techniques and instruments and carried out experimental measurements; CD and JW administered and supervised research operations at UCC; SJB. supervised research operations at OU and provided theoretical guidance. CC and JCD. developed and carried out the comprehensive analysis with key contributions from CD and JW. JW, SJB and JCSD supervised the research project and wrote the paper with key contributions from CD CC and JCD. The manuscript reflects the contributions and ideas of all authors.

**Author Information** Correspondence and requests for materials should be addressed to jcseamusdavis@gmail.com and johnathanward@ucc.ie

**References**

1 P. W. Anderson, Through the glass lightly. *Science* **267**, 1615-1616 (1995).

2 L. Berthier, G. Biroli, J.-P. Bouchaud, L. Cipelletti, W. van Saarloos, *Dynamical Heterogeneities in Glasses, Colloids and Granular Media* (Oxford Univ. Press, 2011).

3 L. Berthier, Dynamic Heterogeneity in Amorphous Materials. *Physics* **4**, 42 (2011).

4 P. Charbonneau *et al*., *Spin Glass Theory and Far Beyond*, (World Scientific, 2023).

5 M. D. Ediger, C. A. Angell, S. R. Nagel, Supercooled liquid and glasses. *J. Phys. Chem.* **100**(31), 13200-13212 (1996).

6 G. Tarjus, S. A. Kivelson, Z. Nussinov, P. Viot, The frustration-based approach of supercooled liquids and the glass transition: a review and critical assessment. *J. Phys.: Condens. Matter* **17**, R1173 (2005).

7 A. Cavagna, Supercooled liquids for pedestrians. *Phys. Reports* **467**, 51-124 (2009).

8 F. Arceri *et al*., *Glasses and Aging: A Statistical Mechanics Perspective* (Encyclopedia of Complexity and Systems Science, 2022).

9 G. Adam, J. Gibbs, On the Temperature Dependence of Cooperative Relaxation Properties in Glass-Forming Liquids. *J. Chem. Phys.* **43** 139 (1965).

10 M. Goldstein, Viscous Liquid and the Glass Transition. *J. Chem. Phys.* **51** 3728 (1969).

11 B. W. H. van Beest, G. J. Kramer, R. A. van Santen, Force fields for silicas and aluminophosphates based on ab initio calculations. *Phys. Rev. Lett.* **64** 1955 (1990).

12 F. Sciortino, Potential energy landscape description of supercooled liquids and glasses. *J. Stat. Mech.* **2005**, P05015 (2005).

13 C. Dalle-Ferrier *et al*., Spatial correlations in the dynamics of glassforming liquids: Experimental determination of their temperature dependence. *Phys. Rev. E* **76**, 041510 (2007).

14 T. Kawasaki, H. Tanaka, Apparent violation of the fluctuation-dissipation theorem due to dynamic heterogeneity in a model glass-forming liquid. *Phys. Rev. Lett.* **102**, 185701 (2009).

15 A. Vila-Costa *et al*., Emergence of equilibrated liquid regions within the glass. *Nat. Phys*. **19**, 114-119 (2023).

16 G. Jung, G. Biroli, L. Berthier, Dynamic heterogeneity at the experimental glass transition predicted by transferable machine learning. *Phys. Rev. B* **109**, 064205 (2024).

17 J. G. Rau, M. J. P. Gingras, Spin slush in an extended spin ice model. *Nat. Commun* **7**, 12234 (2016).

18 M. Udagawa, L. Jaubert, C. Castelnovo, R. Moessner, Out-of-equilibrium dynamics and extended textures of topological defects in spin ice. *Phys. Rev. B* **94**, 104416 (2016).

19 J. N. Hallén, S. A. Grigera, D. A. Tennant, C. Castelnovo, R. Moessner, Dynamical fractal and anomalous noise in a clean magnetic crystal. *Science* **378**, 1218-1221 (2022).

20 V. Raban, L. Berthier, P. C. W. Holdsworth, Violation of the fluctuation-dissipation theorem and effective temperatures in spin ice. *Phys. Rev. B* **105**, 134431 (2022).

21 A. M. Samarakoon *et al*., Structural magnetic glassiness in the spin ice Dy2Ti2O7. *Phys. Rev. Res.* **4**, 033159 (2022).

22 B. C. den Hertog, M. J. P. Gingras, Dipolar Interactions and Origin of Spin Ice in Ising Pyrochlore Magnets. *Phys. Rev. Lett.* **84**, 3430-3433 (2000).

23 A. P. Ramirez, A. Hayashi, R. J. Cava, R. Siddharthan, B. S. Shastry, Zero-point entropy in spin ice. *Nature* **399**, 33-335 (1999).

24 R. G. Melko, M. J. P. Gingras, *Phys. Rev. Lett*. **87**, 067203 (2001).

25 R. G. Melko, M. J. P. Gingras, Monte Carlo studies of the dipolar spin ice model. *J. Phys.: Condens. Matter* **16**, R1277 (2004).

26 I. A. Ryzhkin, Magnetic relaxation in rare-earth oxide pyrochlores. *J. Exp. Theor. Phys*. **101**, 481–486 (2005).

27 C. Castelnovo, R. Moessner, S. L. Sondhi, Magnetic monopoles in spin-ice. *Nature* **451**, 42-45 (2008).

28 C. Castelnovo, R. Moessner, S. L. Sondhi, Spin-ice, Fractionalization, and Topological Order. *Annu. Rev. Condens. Matter Phys.* **3**, 35-55 (2012).

29 E. R. Kassner *et al*., Supercooled spin liquid state in the frustrated pyrochlore $Dy_2Ti_2O_7$. *Proc. Natl Acad. Sci. U.S.A.* **112**, 8549 (2015).

30 S. Havriliak, S. Negami, A complex plane representation of dielectric and mechanical relaxation processes in some polymers. *Polymer* **8**, 161 (1967).

31 F. K. K. Kirschner, F. Flicker, A. Yacoby, N. Y. Yao, S. J. Blundell, Proposal for the detection of magnetic monopoles in spin-ice via nanoscale magnetometry. *Phys. Rev. B* **97**, 140402 (2018).

32 R. Dusad *et al*., Magnetic monopoles noise. *Nature* **571**, 234-239 (2019).

33 A. M. Samarakoon *et al*., Anomalous magnetic noise in an imperfectly flat landscape in the topological magnet Dy2Ti2O7. *Proc. Natl Acad. Sci. U.S.A.* **119**, e2117453119 (2022).

34 C.-C. Hsu *et al*., Dichotomous Dynamics of Magnetic Monopole Fluids *Proc. Natl Acad. Sci. U.S.A.* **121** (21) e2320384121.

35 S. Franz, G. Parisi, On non-linear susceptibility in supercooled liquids. *J. Phys.: Condens. Matter* **12** 6335 (2000).

36 C. Toninelli, G. Biroli, D. S. Fisher, Cooperative Behavior of Kinetically Constrained Lattice Gas Models of Glassy Dynamics. *J. Stat. Phys.* **120**:167 (2005).

37 L. Berthier *et al*., Spontaneous and induced dynamic fluctuations in glass formers. I. General results and dependence on ensemble and dynamics. *J. Chem. Phys*. **126**:184503 (2007a).

38 C. Bennemann *et al*., Investigating the influence of different thermodynamic paths on the structural relaxation in a glass-forming polymer melt. *J. Phys.: Condens. Matter* **11** 2179 (1999).

39 C. Donati, S. C. Glotzer, P. H. Poole, Growing Spatial Correlations of Particle Displacements in a Simulated Liquid on Cooling toward the Glass Transition. *Phys. Rev. Lett*. **82**, 5064 (1999).

40 J. P. Garrahan, D. Chandler, Geometrical Explanation and Scaling of Dynamical Heterogeneities in Glass Forming Systems. *Phys. Rev. Lett.* **89**, 035704 (2002).

41 C. Toninelli *et al.,* Dynamical susceptibility of glass formers: Contrasting the predictions of theoretical scenarios. *Phys. Rev. E* **71**, 041505 (2005).

42 E. R. Weeks, J. C. Crocker, D. A. Weitz, Short- and long-range correlated motion observed in colloidal glasses and liquids. *J. Phys.: Condens. Matter* **19** 205131 (2007).

43 O. Dauchot, G. Marty, G. Biroli, G. Dynamical Heterogeneity Close to the Jamming Transition in a Sheared Granular Material. *Phys. Rev. Lett.* **95**, 265701 (2005).

44 A. S. Keys *et al.,* Measurement of growing dynamical length scales and prediction of the jamming transition in a granular material. *Nat. Phys.* **3** 260-264 (2007).

45 L. Berthier *et al.,* Direct Experimental Evidence of a Growing Length Scale Accompanying the Glass Transition. *Science* **310**, 1797 (2005).

46 L. Berthier *et al.,* Spontaneous and induced dynamic fluctuations in glass formers. I. General results and dependence on ensemble and dynamics. *J. Chem. Phys*. **126**, 184503 (2007b).

47 R. Siddharthan, B. S. Shastry, A. P. Ramirez, Spin ordering and partial ordering in holmium titanate and related systems. *Phys. Rev. B* **63**, 184412 (2001).

48 N. Lačević *et al*., Spatially heterogeneous dynamics investigated via a time-dependent four-point density correlation function. *J. Chem. Phys.* **119**, 7372–7387 (2003).

49 L. Berthier, Time and length scales in supercooled liquids. *Phys. Rev. E* **69**, 020201(R) (2004).

50 L. Berthier, G. Biroli, Theoretical perspective on the glass transition and amorphous materials. *Rev. Mod. Phys*. **83**, 587 (2011).

**Supplementary Information for**

# Discovery of Dynamical Heterogeneity in a Supercooled Magnetic Monopole Fluid

## (I) Magnetic Monopole Dynamics in $Dy_2Ti_2O_7$

### Spin Ice Monopoles

The paradigm of emergent magnetic monopoles in spin ice has been comprehensively successful over decades in explaining the experimentally observed dynamics and magnetic properties in dysprosium titanate (*51*). In such spin-ice compounds, e.g., $Dy_2Ti_2O_7$ and $Ho_2Ti_2O_7$, the lowest energy magnetic excitations are emergent magnetic charges (monopoles). Each $Dy^{3+}$ or $Ho^{3+}$ magnetic ion occupies a vertex of the corner-sharing tetrahedral sublattice and exhibits only two magnetic states with dipole moments $\mu \approx 10\mu_B$, pointing either towards or away from the center of each tetrahedron. Moreover, the lowest energy configuration of each tetrahedron is constrained by the dipolar-spin-ice Hamiltonian to have two spins pointing in and two pointing out (2-in:2-out), while the higher energy excitations are the effective magnetic charges ($+m$ for 1-in:3-out and $-m$ for 3-in:1-out) that are in some degree mobile. The magnetization dynamics *(52-59)* of both $Dy_2Ti_2O_7$ and $Ho_2Ti_2O_7$ are now widely viewed as due the correlated transport characteristics of emergent magnetic monopoles. On this basis the existence of a fluid of emergent magnetic monopoles is well attested. It is widely theorized that monopole dynamics continues even down to temperatures approaching 100 mK (*21,60-62*)

## (II) Susceptibility and Relaxation Time Studies of $Dy_2Ti_2O_7$

The magnetic susceptibility $\chi(\omega,T) = \chi'(\omega,T) - i\chi''(\omega,T)$ of $Dy_2Ti_2O_7$ is known empirically with high precision *(58-60,63-70)*, as is the fact that below $T \approx 500$ mK the linear-response relaxation rates in $Dy_2Ti_2O_7$ become ultra-slow *(56,71,72)*. Fig. S1 contains a review of measured linear-response relaxation times of $Dy_2Ti_2O_7$ using different experimental techniques with data from this work included.

A previous high precision study of the magnetic susceptibility of $Dy_2Ti_2O_7$ identified that the frequency-dependence of the magnetic susceptibility is very accurately parametrized by the Havriliak-Negami (HN) equation

$$\chi(\omega, T) = \chi_\infty + \chi_0(T) / \left(1 + \left(i\omega\tau_\chi(T)\right)^{\alpha(T)}\right)^{\gamma(T)} \quad \text{(S1)}$$

Solving for the real and imaginary components of S1

$$\chi' = \chi_\infty + \chi_0 \frac{\cos(\gamma\varphi)}{\left(1+2\left(\omega\tau_\chi\right)^\alpha \cos\left(\frac{\pi\alpha}{2}\right)+\left(\omega\tau_\chi\right)^{2\alpha}\right)^{\gamma/2}} \quad \text{(S2)}$$

$$\chi'' = \chi_0 \frac{\sin(\gamma\varphi)}{\left(1+2\left(\omega\tau_\chi\right)^\alpha \cos\left(\frac{\pi\alpha}{2}\right)+\left(\omega\tau_\chi\right)^{2\alpha}\right)^{\gamma/2}} \quad \text{(S3)}$$

Here $\chi_\infty$ is the real value of $\chi$ in the $\omega \to \infty$ limit and $\tau_\chi$ is the characteristic relaxation time. $\alpha(T)$ and $\gamma(T)$ describe the broadening and asymmetry of the distribution of relaxation times and

$$\varphi = \arctan(\left(\omega\tau_\chi\right)^\alpha \sin\left(\frac{\pi\alpha}{2}\right) / 1 + \left(\omega\tau_\chi\right)^\alpha \cos\left(\frac{\pi\alpha}{2}\right)) \quad \text{(S4)}$$

The HN parametrization of $\chi(\omega, T)$ and the associated broad distribution of relaxation times, rather than a single relaxation time, implies a clear non-Debye character of the susceptibility. Indeed, the $\alpha(T)$ (broadening) and $\gamma(T)$ (asymmetry) terms describe the distribution of relaxation times and themselves evolve through the supercooled regime as previously discovered (*29*). Further, the linear-response relaxation times extracted by fitting directly measured $\chi(\omega, T)$to S1 was demonstrated to be of the form

$$\tau_\chi(T) = A\exp\left(DT_g / (T - T_g)\right) \quad \text{(S5)}$$

where $D = 13.6 \pm 5.0$ is the fragility index of the glass-forming state and $T_g \approx 240$ mK $\pm$ 30 mK. This is the Vogel–Tammann–Fulcher (VTF) form characteristic of a supercooled glass-forming molecular liquids. Hence, these forms for the susceptibility $\chi(\omega, T)$ and the relaxation time $\tau_\chi(T)$ indicate the existence of a supercooled monopole liquid in $Dy_2Ti_2O_7$, a deduction that is consistent with the empirical $\chi(\omega, T)$ and $\tau_\chi(T)$ reported by virtually all studies (Fig. S1).

## (III) Combined Monopole Noise Spectrometer and AC Susceptometer

### Design

Our monopole noise spectrometer assembly is shown schematically in Fig. S2. The sample holder is a hollow Macor cylinder onto which two persistent superconducting coils (signal pick up and field-cancellation coil) wound with opposite chirality are connected in-series with the input coil of the Quantum Design Model 550 SQUID. The SQUID couples a ∼ *1* μH input coil into the ∼ *100* pH SQUID coil, while maintaining a nominal critical current of ∼ *10* μA. A cylindrical superconductive 'drive' coil for applying $\mu T$ magnetic fields to the sample surrounds the pickup and astatic coils. The experiment is mounted at the mixing chamber plate of a dilution refrigerator. To expel and shield external magnetic fields, the SQUID is shielded within its own Niobium shield, this stage is surrounded by an additional outer Niobium cylindrical shield which is in turn enclosed in a larger cylindrical mu-metal shield. The spectrometer is mounted on the mixing chamber plate of a low-vibration cryogen-free dilution refrigerator that is vibrationally isolated and enclosed inside an acoustic isolation chamber. The refrigerator reaches a base temperature of 12 mK.

### Flux Noise Acquisition

The time-sequence of the magnetic flux generated by the sample $\Phi_p(t)$ is extracted using the inductances of the pickup coil $L_p$ and input coil $L_i$, and $\mathcal{M}_i$ the mutual inductance to SQUID

$$\Phi_p(t,T) \equiv \Phi(t,T)/\left(\frac{\mathcal{M}_i}{2L_p+L_i}\right) \equiv \beta V(t,T)/G \tag{S6}$$

Where $\beta \equiv \left(\frac{\mathcal{M}_i}{2L_p+L_i}\right)^{-1} = 185.3$ derived from $L_p = 0.71$ μH ; $L_i = 1.74\ \mu H; \mathcal{M}_i = 1.1 \times 10^{-8}\ \Phi_0/\mu A$ as set by the coil design. Using a SR560 Voltage Preamplifier, the signal is amplified and filtered by a low pass filter with a cutoff frequency $f_{LP}$ of 3 kHz, above which the SQUID is bandwidth limited. For temperatures above 600 mK, an additional high pass filter is added with cutoff $f_{HP}$ of 0.03 Hz. The filtered SQUID output voltage $V$ is recorded with 10 microsecond resolution for a total time of 1000 seconds.

**Magnetic Susceptibility Data Acquisition**

AC susceptibility measurements use a SR830 lock-in amplifier to measure the in-phase and out-of-phase components of the voltage output of the SQUID. An AC magnetic field $H(\omega)$ is synthesized by the lock-in amplifier by passing the Sine Output voltage signal (10 $\mathrm{mV_{RMS}}$) through a 20 kΩ resistor and RF filter before entering the solenoid drive coil (Fig. S2). The response of the $Dy_2Ti_2O_7$ sample is measured by the SQUID and fed into the lock-in amplifier. At each temperature setpoint, four frequency ranges are recorded: $0.1, 0.3, \dots, 0.9$ Hz; $1, 2, \dots, 10$ Hz; $11, 21, \dots, 101$ Hz ; $100, 200, 500, 1000$ and $2000$ Hz . The time constant is chosen to be $\tau_{LI} \geq 3(1/f_{\min})$ for the respective frequency ranges. The sensitivity of the lock-in amplifier is set to 20 mV/nA for $T < 600$ mK and 50 mV/nA for $T \geq 600$ mK.

## (IV) Monopole Noise and AC Susceptibility Analysis

**Noise Analysis**

The magnetization is related to the output voltage of the SQUID as

$$V(t,T) = \Phi_p(t,T)\, G/\beta = \frac{M(t,T)}{C_0} \tag{S7}$$

where $C_0 \equiv \left(\frac{\Phi_0}{\beta NAF}\right) = 2.1 \times 10^{-9}\ \mathrm{JT^{-1}V^{-1}m^{-3}}$ is calibrated accurately for our experimental geometry. The time-sequences of magnetization fluctuations are recorded from $V(t)$ for each temperature $T$. The power spectral density of magnetization noise $S_M(\omega, T)$ is derived using

$$S_M(\omega, T) \equiv \lim_{\mathcal{T}\to\infty} \frac{1}{\pi\mathcal{T}} \left| \int_{-\frac{\mathcal{T}}{2}}^{\frac{\mathcal{T}}{2}} M(t) e^{-i\omega t} dt \right|^2 \tag{S8}$$

The complete frequency and temperature dependence of the magnetization noise spectral density is shown in Fig. S3.

### Magnetic Susceptibility Analysis

To calculate the AC Susceptibility, it is convenient to first define a pre-factor $F_1 = C_\chi(2L_p + L_i)/\mathcal{M}_i$ for converting the SQUID output voltage to magnetic flux in the pickup coil. $C_\chi = 0.0073\ \mathrm{V}/\Phi_0$ is a value intrinsic to the SQUID electronics, while $L_p = 0.71\ \mu\mathrm{H}$ and input coil $L_i = 1.74\ \mu\mathrm{H}$ represent the inductances of the pickup coil and input coil respectively. $\mathcal{M}_i = 1.1 \times 10^{-8}\ \Phi_0/\mu\mathrm{A}$ represents the mutual inductance of the SQUID circuitry (Fig. S2). To convert flux to $B$-field, a second pre-factor $F_2 = \Phi_0/N_{coil}A_{coil}F$ is used. $N_{coil} = 16$ is the total number of turns in the pickup coil, $A_{coil} = 3.843 \times 10^{-6}\ \mathrm{m}^2$ is the pickup coil cross-sectional area, $F = 0.57$ is the filling factor. At each frequency 10 in-phase (X) and out-of-phase (Y) voltage values are collected from the Lock-In, from which average values $V_x$ and $V_y$ are calculated. Quantitatively accurate real and imaginary magnetic susceptibilities are then found using

$$\chi'(\omega, T) = \frac{V_x(\omega,T)}{\mu_0 H_{mod}}\left(\frac{1}{F_1 F_2}\right) \quad \text{(S9)}$$

$$\chi''(\omega, T) = \frac{-V_y(\omega,T)}{\mu_0 H_{mod}}\left(\frac{1}{F_1 F_2}\right) \quad \text{(S10)}$$

$\chi'$ and $\chi''$ are fitted to the HN equations S2 and S3 respectively and presented on both a linear and logarithmic scale in Fig. S4.

## (V) Ergodicity from Fluctuation-Dissipation Theorem Analysis

### Examining Ergodicity of the Monopole Fluid

If the Fluctuation-Dissipation Theorem (FDT) is obeyed for $Dy_2Ti_2O_7$, the magnetization noise $S_M(\omega, T)$ would be directly related to the imaginary AC susceptibility $\chi''$ by

$$S_M(\omega, T) = \frac{2k_B T}{\omega\pi\upsilon\mu_0}\chi''(\omega, T) \quad \text{(S11)}$$

wherein SI units are used throughout so that $\chi''(\omega, T)$ is unitless. Using measured $S_M(\omega, T)$ and $\chi''(\omega, T)$, the left-hand side of S11 is plotted against the right-hand side for frequencies in the range 0.3 − 2000 Hz (Fig. S5). To improve the reliability of the low temperature noise data where the signal to noise ratio is lowest, each $S_M(\omega, T)$ (which has the contribution from the empty coil subtracted) curve is averaged over 20 second segments, meaning the error bars (~1%) are not overlapping with the noise floor even for the lowest magnitude noise data. Each temperature, differentiated by color in Fig. 1F in the main text, contains several points on the curve corresponding to the frequencies used in the experiment. To quantify the validity of the FDT, a ratio $X(\omega, T)$ is defined as

$$X(\omega, T) = \frac{2k_\mathrm{B}T}{\omega\pi\upsilon\mu_0}\frac{\chi"(\omega,T)}{S_M(\omega,T)} \tag{S12}$$

Where $X \approx 1$, the FDT is obeyed while $X < 1$ indicates a violation of FDT due to a loss of ergodicity of the system and the presence of excess noise. To show the temperature evolution, $\overline{X}$(T) is defined to be $X(\omega, T)$ averaged over all experimental frequencies. $\overline{X}$(T) is shown in Fig. 4Dii in the main text.

## (VI) Analysis of Time-Resolved Monopole Noise

### Flux at Pickup Coil from SQUID Output

The SQUID output voltage signal $V(t, T)$ is recorded with 10 μs precision. $V(t, T)$ is calibrated by the design of the circuit (Fig. S2) to accurately measure the flux produced by the $Dy_2Ti_2O_7$ crystal as it threads the pickup coil $\Phi_p(t, T)$ as in Eqn. S7. A typical $\Phi_p(t, T)$ signal is shown in Fig. 1B in the main text.

### Magnetic Monopole Current

The monopole current $J(t, T)$ is in principle related to the flux $\Phi_p(t, T)$ by

$$J(t, T) \equiv \dot{\Phi}_p(t, T)/\mu_0 \tag{S13}$$

When calculating the time derivative of a noisy $\phi_p(t, T)$ signal, an 80 μs boxcar average is first applied to suppress artifacts that may arise from numerical differentiation. The derivative $\dot{\Phi}_p(t, T)$ is calculated using the Central Difference Method:

$$\dot{\Phi}_p(t, T) = \frac{\Phi_p(t+\Delta t, T) - \Phi_p(t-\Delta t, T)}{2\Delta t} \tag{S14}$$

Using S13 the current $J(t, T)$ is calculated. In this analysis, only the magnitude of current noise $|J(t, T)|$ as no net current is observed. In particular, the distribution of occurrence rate $r_{|J|}$, is calculated by considering the number $\eta(|J|)$ of times a given current magnitude $|J|$ occurs in a fixed time interval $P$: $r_{|J|} = \eta(|J|)/P$. Further analysis examines the mean of monopole current magnitudes $|J|$ versus temperature $T$. Results of the magnitude of monopole current $|J(t, T)|$ are presented in Fig. 2 in the main text. Two types of monopole current occur within this current distribution: rearranging S13 the relation which directly relates $J(t)$ to changes in the flux $\Phi_p(t)$ is

$$\mu_0 \int_{t_i}^{t_f} J(t', T)\, dt' = \Phi_p(t_f, T) - \Phi_p(t_i, T) \tag{S15}$$

This means that intense current bursts existing over extended time periods produce excursions in $\Phi_p(t, T)$ far larger than those generated by conventional monopole noise. This effect is seen directly in histograms of $|\Phi_p(t, T)|$ as show in Fig. S6.

**Energetics: Continuous Distribution of Energies**

To understand the energy scales of the monopole phenomena, the relation

$$\varepsilon(t, T) \equiv \Phi_p^2(t, T)/2L_p \tag{S16}$$

is used. In general, a flux change $\Phi$ in a persistent superconducting coil of inductance $L$ produces a supercurrent given by $I = \Phi/L$ : the coil then stores energy $\varepsilon = \Phi^2/2L$ . Hence in our studies, where the only source of energy is the magnetization dynamics of the sample, the flux produced by the $Dy_2Ti_2O_7$ sample at the pickup coil, $\Phi_p$, represents a monopole current event with energy $\varepsilon_p = \Phi_p^2/2L_p$. Fig. 3A in the main text shows that monopole current bursts, which are large collective increases in the flux always followed by a collective reversal, typically occur on timescales of order ~1 ms. The square of the flux noise signal $\Phi_p^2$ is averaged in an 80 μs window for consistency with the current analysis. The continuous $\Phi_p^2(t, T)$ signal (Fig. S7) is converted to energy using S16. For reference, the noise picked up purely by the circuitry (no $Dy_2Ti_2O_7$ sample) is shown in black in Fig. S7. The distribution of the occurrence rate $r(\varepsilon, T)$ of events with energy $\varepsilon$ is calculated by considering the number $m(\varepsilon)$ of times a given energy $\varepsilon$ occurs in the continuous energy signal within a fixed time interval $P$: $r(\varepsilon) = m(\varepsilon)/P$. The striking emergence (see Movie and Audio S1) of a second gaussian distribution in the range 250 mK $\lesssim T \lesssim$ 1500 mK, corresponding to the emergence of current bursts in the $\Phi_p^2$ signal, prompts further analysis. To do so, a given $r(\varepsilon, T)$ distribution is fit to a bi-modal model, where the overall distribution is represented by the sum of two unique gaussian functions

$$\varepsilon_M + \varepsilon_B = A_M \exp\left(-\frac{(\varepsilon - \overline{\varepsilon_M})^2}{2\sigma_M^2}\right) + A_B \exp\left(-\frac{(\varepsilon - \overline{\varepsilon_B})^2}{2\sigma_B^2}\right) \quad \text{(S17)}$$

Here subscript M denotes the noise produced by conventional monopole noise and subscript B denotes the noise produced by transient bursts of monopole current. In the cases where this model fails (i.e., one of the distributions goes to zero, or the two gaussians are almost completely overlapping), it implies that the current bursts are no longer present in the signal. Results of the analysis of the continuous distribution of energies are presented in Fig. 3 in the main text.

## (VII) Monopole Noise Power Law

The magnetization noise floor, as measured using an empty pickup coil, is subtracted from the measured $Dy_2Ti_2O_7$ magnetization noise at each temperature. The resulting magnetization noise spectrum $S_M$ reveals the true contribution to the magnetization signal from the monopoles. $S_M$ is fitted using a least-squares method to the standard equation

$$S_M(\omega, T) = \frac{\sigma_M^2(T)\tau_N(T)}{\left(1 + \left(\omega\tau_N(T)\right)^{b(T)}\right)} \quad \text{(S18)}$$

in the frequency range $0.05 - 10{,}000$ rad/s. For optimal fitting, only data two times greater than the noise floor are included in the fit. The power law exponent $b(T)$, relaxation time $\tau_N(T)$ and magnetization variance $\sigma_M^2(T)$ are free parameters of the fit. Within our 1000s acquisition, it is possible to access very slow relaxation times $\tau_N(T)$ on the order of tens of seconds (Fig. S1). The quality of fit is indicated by the inset of Fig. S8. Fig. 4Diii in the main text shows the temperature dependence of the monopole noise power law $b(T)$; a sharp decrease from the predicted $b = 1.5$ towards $b = 1$ is seen in the $T \rightarrow 0$ limit.

## (VIII) Autocorrelation Function

The autocorrelation function $F_{\Phi}(\tau, T)$ is defined $F_{\Phi}(\tau, T) \equiv N_F(T)\langle \Phi_p(t, T)\Phi_p(t + \tau, T)\rangle_t$. $F_{\Phi}(\tau, T)$ is calculated from the discrete flux signal $\Phi_p(t, T)$, in the interval $P = 1000$ s, by:

$$F_{\Phi}(\tau, T) = N_F(T)\frac{1}{P-\tau}\sum_{t=0}^{P-\tau}\Phi_p(t, T)\Phi_p(t + \tau, T) \tag{S19}$$

where $N_F = P/(\sum_{t=0}^{L}\Phi_p(t, T)\Phi_p(t, T))$ normalizes $F_{\Phi}(\tau, T)$ ensuring $F_{\Phi}(\tau = 0) = 1$ . $F_{\Phi}(\tau, T)$ is calculated from each $\Phi_p(t, T)$ and interpolated (linear) at $\Delta_T$= 10 mK intervals to create a smooth surface. The evolution of $F_{\Phi}(\tau, T)$ from the monopole fluid regime to the deeply supercooled regime is shown in Fig. S9. The evolution of the relaxation time is clearly identified and correlated monopole motion increases dramatically as temperature falls.

## (IX) Four-Point Dynamical Susceptibility: Theory and Experiment

### Deriving Dynamical Susceptibility from the Autocorrelation Function

Numerical works on supercooled glass-forming liquids show the emergence of spatially heterogeneous dynamics *(73)* upon cooling into the supercooled phase. Subsequent studies have attempted to quantify this fact by probing four-point dynamic susceptibility

$$\chi_4(\tau, T) \equiv N\langle \delta C^2(t, \tau, T)\rangle_t \tag{S20}$$

which measures the spatiotemporal correlations of fluctuations about the average. $C_A(t, \tau, T) = A(t, T)A(t + \tau, T)$ represents the two-point correlation of an instantaneous fluctuation of a mean-zero local observable $A(\vec{r}, t, T)$. Here $\langle \delta C^2\rangle = \langle C_A^2\rangle - \langle C_A\rangle^2$ is the ensemble-averaged fluctuation of $C_A^2(t, \tau, T)$ about its average, and $N$ is the number of particles. The normalized correlation function $F_A(\tau, T)$ is equivalent to the average of $C_A(t, \tau, T)$ over the time $t$ of the observation: $F_A(\tau, T) \equiv \langle C_A(t, \tau, T)\rangle_t / \langle C_A(t, 0, T)\rangle_t$.

As discussed in the main text, this spatiotemporal information is not currently available for molecular glass forming liquids. Alternatively, using an approach based on fluctuation-dissipation theorem, the dynamic susceptibility can be defined from the time-dependent correlator $F(\tau, T)$. First, the response function $\chi_T(\tau, T)$ is defined as the response of $F(\tau, T)$ to temperature variations

$$\chi_T(\tau, T) = \frac{\partial F(\tau, T)}{\partial T} \tag{S21}$$

This equation also holds in the frequency domain $\chi_T(\omega, T) = \partial \tilde{F}(\omega)/\partial T$ where $\tilde{F}(\omega)$ can be the dielectric susceptibility. In a molecular liquid, the fluctuation-dissipation relation

$$k_B T^2 \chi_T(\tau, T) = N\langle \delta C(t, \tau, T) \delta H(t, 0, T)\rangle \tag{S22}$$

can be established, where $k_B$ is the Boltzmann constant, $\delta H(t, 0, T)$ the fluctuating enthalpy per particle and $\delta C(t, \tau, T)$ is the instantaneous value of the correlation function $F(\tau, T)$. Importantly here, $C(t, \tau, T)$ and $H(t, \tau, T)$ are sums over local contributions (*74*)

$$C(t, T) = \frac{1}{V} \int d^3\vec{r}\ c(\vec{r}, t, T) \tag{S23}$$

$$H(t, \tau, T) = \frac{\sqrt{k_B c_p(T)} T}{V} \int d^3\vec{r}\ h(\vec{r}, t, T) \tag{S24}$$

where $V$ is the volume of the sample. From thermodynamics, the specific heat at constant pressure $c_p(T)$ here sets the scale of the enthalpy fluctuations $\langle \delta H(t, 0, T)^2\rangle_t = N k_B c_p(T) T^2$. Using translational invariance, Eqn. S22 is rewritten as

$$\sqrt{\frac{k_B}{c_p(T)}} T \chi_T(\tau, T) = \int d^3\vec{r}\ \langle \delta c(\vec{r}, t, \tau, T) \delta h(0, t, 0, T)\rangle_t \tag{S25}$$

where $\rho = N/V$ is the density. In the same way the four-point correlation function is the variance of the two-point correlation function $\langle \delta C^2(t, \tau, T)\rangle_t \propto \int d^3\vec{r}\ \langle \delta c(\vec{r}, t, \tau, T) \delta c(r, t, 0, T)\rangle_t$, the space integral of the three-point correlation is the covariance of the dynamic correlation with energy fluctuations: $\langle \delta C(t, \tau, T) \delta H(t, 0, T)\rangle_t \propto \int d^3\vec{r}\ \langle \delta c(\vec{r}, t, \tau, T) \delta h(0, t, 0, T)\rangle_t$. This powerful relationship shows that $\chi_T(\tau, T)$, and therefore the flux signal $\Phi_p(t,T)$ itself, allows one to directly probe the spatial correlations between local fluctuations of the dynamics and that of the enthalpy.

It is important to note here that the flux is a valid choice of measurable quantity to observe these effects despite having no spatial coordinate. Using a volume averaged quantity like $\Phi_p$ (directly proportional to magnetization $M(t)$ ) to calculate $C(t,\tau,T)$ represents approximately taking the $q=0$ Fourier component of the integral $C(t,\tau,T) = \frac{1}{V}\int d^3\vec{r}\, c(\vec{r},t,\tau,T) = \frac{1}{V}\int d^3\vec{r}\, \Phi(\vec{r},t+\tau,T)\,\Phi(\vec{r},t,T) = \frac{1}{V(2\pi)^3}\int d^3\vec{q}\, \Phi(\vec{q},t+\tau,T)\,\Phi(-\vec{q},t,T)$. In this sense, the spatial correlation is not directly in the measurable quantity but can be inferred through the relationship between $\chi_T(\tau,T)$ and $\chi_4(\tau,T)$. This approach has been successfully demonstrated in experiments on colloidal hard spheres and glass-forming glycerol.

Specifically, $\chi_T$ can be related to $\chi_4$ using the Cauchy-Schwartz inequality:

$$\langle \delta H(t,0,T)\delta C(t,\tau,T)\rangle_t^{\,2} \leq \langle \delta H(t,0,T)^2\rangle_t \langle \delta C(t,\tau,T)^2\rangle_t \tag{S26}$$

Using Eqn. S22 to substitute for $\chi_T(\tau,T)$ on the left-hand side and Eqn. S20 to substitute for $\chi_4(\tau,T)$ on the right-hand side

$$\frac{1}{N^2}(k_B T^2)^2[\chi_T(\tau,T)]^2 \leq \frac{1}{N}(k_B c_p(T) T^2)\frac{1}{N}\chi_4(\tau,T) \tag{S27}$$

$$\chi_4(\tau,T) \geq \frac{k_B T^2}{c_p(T)}[\chi_T(\tau,T)]^2 \tag{S28}$$

Now, the experimentally accessible $\chi_T(\tau,T)$ can be used to tightly bound the dynamic susceptibility $\chi_4(\tau,T)$. In general, to equate the two, all fluctuating quantities $\Delta_i$ in the system should be considered (*75*). $\chi_4(\tau,T)$ is the sum of all contributions from fluctuations: $\chi_4(\tau,T) = \sum k_i(T)\,\chi_{\Delta_i}(\tau,T)$. Here $k_i(T)$ represents the collection of pre-factors relevant to each fluctuation. To first order, the temperature and monopole density $\rho_{\mathrm{mono}}$ fluctuations are relevant in the supercooled monopole fluid of $Dy_2Ti_2O_7$,

$$\chi_4(\tau,T) \approx \frac{k_B T^2}{c_p(T)}[\chi_T(\tau,T)]^2 + \rho k_B T \kappa_T \rho_{\mathrm{mono}}^2 \left[\chi_{\rho_{mono}}(\tau,T)\right]^2 \tag{S29}$$

where $\kappa_T$ is the isothermal compressibility and $\rho$ the density of $Dy_2Ti_2O_7$. Studies on fragile liquids (*76*), which is the case of the monopole fluid in $Dy_2Ti_2O_7$, reveal that the density term is negligible, therefore $\chi_T$ itself provides a good approximation of $\chi_4$.

$$\chi_4(\tau, T) \approx \frac{k_b T^2}{c_p(T)} [\chi_T(\tau, T)]^2 \quad \text{(S30)}$$

Because the FDT begins to be violated in the supercooled monopole fluid at $T \approx 500$ mK, $\chi_4(\tau, T)$ is only considered above this temperature.

**Response Function Analysis of Supercooled Monopole Fluids**

The dynamical susceptibility described above has been derived, simulated, and experimentally measured for conventional molecular glass forming liquids. However, this approach need not be restricted to conventional glass formers but may also be relevant to spin glasses. It has been shown that the growth of the dynamical susceptibility derived from Eqn. S30 reveals unique dynamical length scales that are independent of the choice of dynamics and, most importantly, independent of the choice of ensemble. The requirements to apply the above concepts are the validity of the fluctuation-dissipation theorem and that the system must obey the general thermodynamic relation that the temperature derivative of a global dynamical correlator is related to cross-correlations (via fluctuation–dissipation) between local dynamical fluctuations and energy (enthalpy). The monopole fluid can be well interpreted within the grand canonical ensemble (*77*) and remains ergodic at $T \geq 500$ mK, hence the dynamical susceptibility of the monopole fluid may be described as above.

The use of flux noise $\Phi(t)$ measurements allows greatly increased sensitivity at lowest frequencies because, when FDT holds, $S_M(\omega, T)/\chi''(\omega, T) \propto T/\omega$, so that in terms of its signal/noise ratio, accessing extremely slow $\omega \to 0$ dynamics using AC susceptibility $\chi''(\omega, T)$ is enormously less efficient than doing so from long-time magnetization noise $\Phi(t)$ measurements.

**Calculating the Dynamical Susceptibility from Noise Data**

The multipoint correlation function $\chi_T(\tau, T)$ is calculated directly from the $F_\Phi(\tau, T)$ surface shown in Fig. S9 using the Central Difference Method

$$\chi_T(\tau, T_i) \equiv \frac{\partial F_\Phi(\tau,T_i)}{\partial T} = \frac{F_\Phi(\tau,T_{i+1})-F_\Phi(\tau,T_{i-1})}{2\Delta_T} \qquad \text{(S31)}$$

$\chi_T(\tau, T)$ is shown across the supercooled regime in Fig. S10. The heat capacity at constant pressure $c_p(T)$ is calculated from the data reported in (*47* pp 5). Using $c_p(T)$ and $\chi_T(\tau, T)$ data as shown in Fig. S10, $\chi_4(\tau, T)$ is calculated using Eqn. S30. Because the low-frequency monopole noise grows exponentially with falling temperature, probing the dynamic susceptibility using noise allows for far greater signal-to-noise ratio than ac susceptibility experiments would allow for.

## (X) Nonlinear Susceptibility Studies of Dynamical Heterogeneity

### Growing Length Scales in Amorphous Materials

In spin glasses, it has been shown (*78*) that amorphous order can be detected only via higher order magnetic susceptibilities, because the four-point susceptibility $\chi_4(t)$ and the growing cooperative volume ,while inaccessible experimentally, are shown theoretically to be connected to the higher order susceptibility (*13*). Higher order susceptibilities have been measured directly in spin glasses and have been extracted indirectly in supercooled liquids. In spin glasses, higher order susceptibilities have been measured directly by non-linear magnetization (*78*). In supercooled liquids, higher order susceptibilities have been extracted from the temperature derivative (as derived in Section IX above) of the linear susceptibility (*45, 79-81*). In all of these models, the non-linear susceptibilities of amorphous materials reveal growing length scales approaching the freezing/glass transition temperature. Beyond susceptibility, the growing length scale associated with dynamical heterogeneity has also been measured in a polymer using multidimensional NMR techniques (*82*).

## (XI) $Dy_2Ti_2O_7$ Samples

### Sample Growth

The single crystal rod-shaped $Dy_2Ti_2O_7$ samples (Fig. S11) are grown by floating zone method. High purity (99.99%) $Dy_2O_3$, and $TiO_2$ are mixed and heated to 1400 °C for 40 hours. The mixture is ground immediately, then heated for 12 hours. The resulting powder is packed into a rod, then sintered at 1400° C for 12 hours. A long piece of the sintered rod is used as a feed rod while a small piece is used as the seed. The crystals are grown in 0.4 MPa oxygen pressure at 4 mm/hour using a two-mirror NEC furnace where the feed and seed rods are counter-rotated at 30 rpm.

**Sample Demagnetization**

The demagnetization factor for a rod-shaped sample with the field applied along the length of the rod is zero for an infinitely long sample. However, real experiments have finite length and therefore a finite demagnetization factor needs to be accounted for. A finite cuboid has a demagnetizing field $B_d$ of

$$\vec{B}_d = -\boldsymbol{N}(\vec{r})\mu_0\vec{M} \tag{S32}$$

where $\boldsymbol{N}(\vec{r})$ is the position dependent demagnetization factor. For a cuboid with dimensions $2a \times 2b \times 2c$, $N(\vec{r})$ is given by *(85)*

$$\boldsymbol{N}_{ii}(\vec{r}) = \frac{1}{4\pi}\sum_{\alpha=\pm1}\sum_{\beta=\pm1}\sum_{\gamma=\pm1}\tan^{-1}(f_i(\alpha x,\beta y,\gamma z)) \tag{S33}$$

With $i = x, y, z$ and the functions $f_i$ are given by

$$f_x(x,y,z) = \frac{(b-y)((c-z)}{(a-x)\sqrt{[(a-x)^2+(b-y)^2+(c-z)^2]}} \tag{S34}$$

$$f_y(x,y,z) = \frac{(a-x)((c-z)}{(b-y)\sqrt{[(a-x)^2+(b-y)^2+(c-z)^2]}} \tag{S35}$$

$$f_z(x,y,z) = \frac{(b-y)((a-x)}{(c-z)\sqrt{[(a-x)^2+(b-y)^2+(c-z)^2]}} \tag{S36}$$

Considering the superconducting pickup coil is wound around the centre of the DTO sample, the greatest contribution to the demagnetizing field will be along the z axis at the centre of the coil, $\boldsymbol{N}_{zz}(0,0,0)$. Using the known geometry of the sample $2a = 0.98$ mm, $2b = 1.31$ mm, $2c = 6.23$ mm, this gives an approximate demagnetization factor of

$$\boldsymbol{N}_{zz(0,0,0)} \sim 0.01 \tag{S37}$$

which was part of the experimental design concept. Therefore, the estimated demagnetization factor is roughly 1%, which is insignificant to observations reported in this work.

**Stuffed Defects**

The floating zone method used here for growing pyrochlore titanate crystals typically produces clean crystals with ~1% stuffing fraction *(84)*. These stuffed sites undoubtedly produce their own dynamics which have not been extensively studied in previous works, while certainly influencing the monopolar dynamics particularly at low temperatures (*85*). At 1.5 K when the current bursts begin to emerge, the monopolar density is at least an order of magnitude greater than the defect density. Further, at this temperature the magnetization noise spectrum is consistent with a fluid of magnetic monopoles as the source of magnetization intrinsic to the sample. It is therefore implausible that the monopole bursts can be attributed to site disorder/stuffing in the crystal. A key future project could be the investigation of the effects of stuffed defect density on the supercooled monopole fluid state of DTO.

**Repeatability**

This sequence of experiments was repeated with three different $Dy_2Ti_2O_7$ samples. Within typical margins due to geometrical effects, all samples produced equivalent phenomenologies (Fig. S12). The reported magnetization noise and susceptibility data show excellent agreement with previously reported data on the material, validating the quality of samples and experimental assembly.

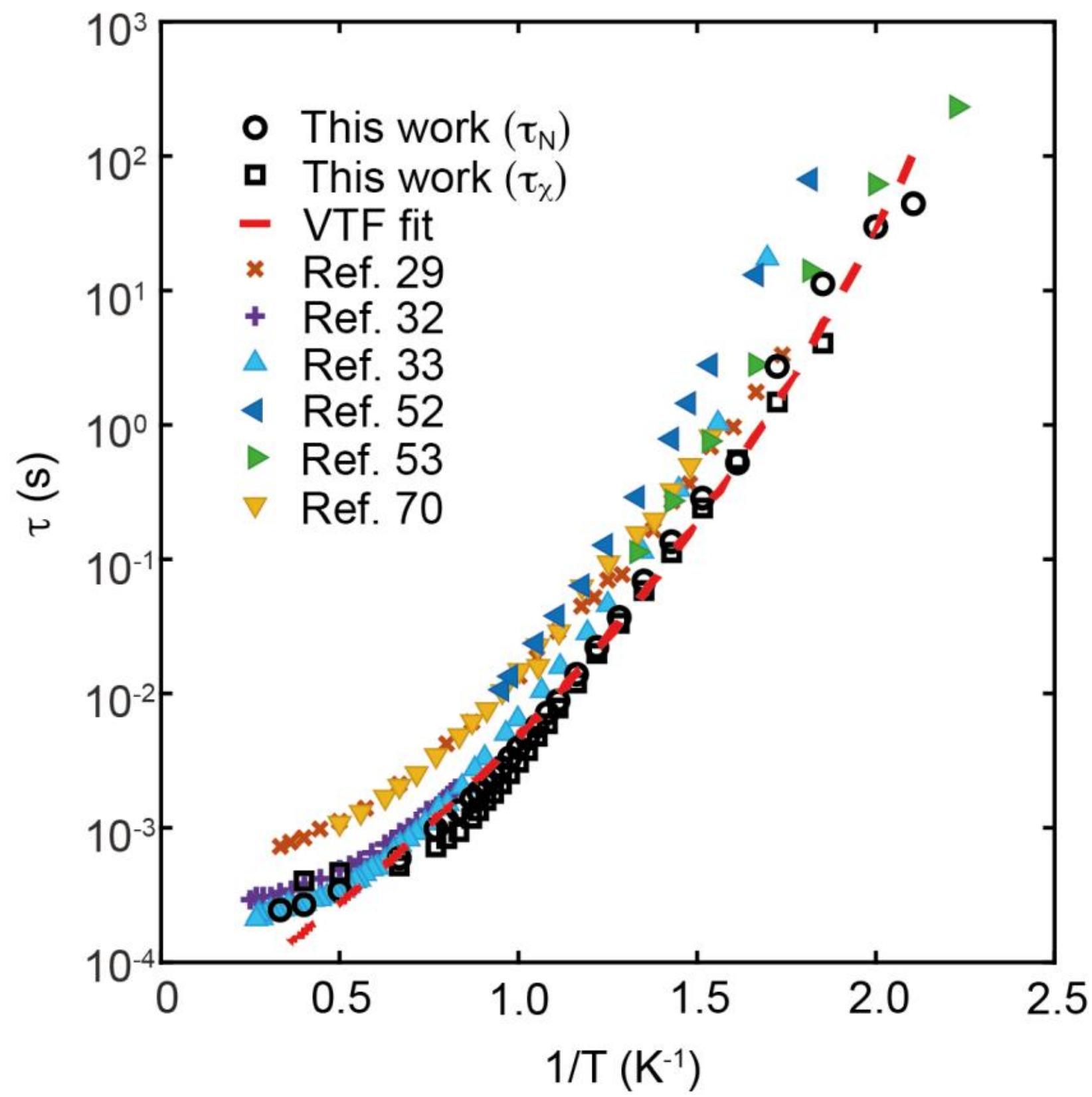

**Fig. S1:** The linear-response relaxation time $\tau$ measured by fitting our magnetization noise $S_M$ (black circles) and AC susceptibility $\chi$" (black squares) is compared to related measurements in the literature (coloured symbols) and found to be consistent with previously reported values. Below $T \approx 500$ mK, $\tau$ becomes inaccessible to linear-response experiments due to its divergent behavior approaching $T_g = 240 \pm 30$ mK. Refs. 32 & 33 derive $\tau$ from noise measurements, while Refs. 29, 52, 53 and 70 derive $\tau$ from AC susceptibility measurements. Deviations in values of $\tau(s)$ between the referenced data sets are attributed to due to different sample geometries and experimental configurations. However, the VTF form of relaxation time is held throughout.

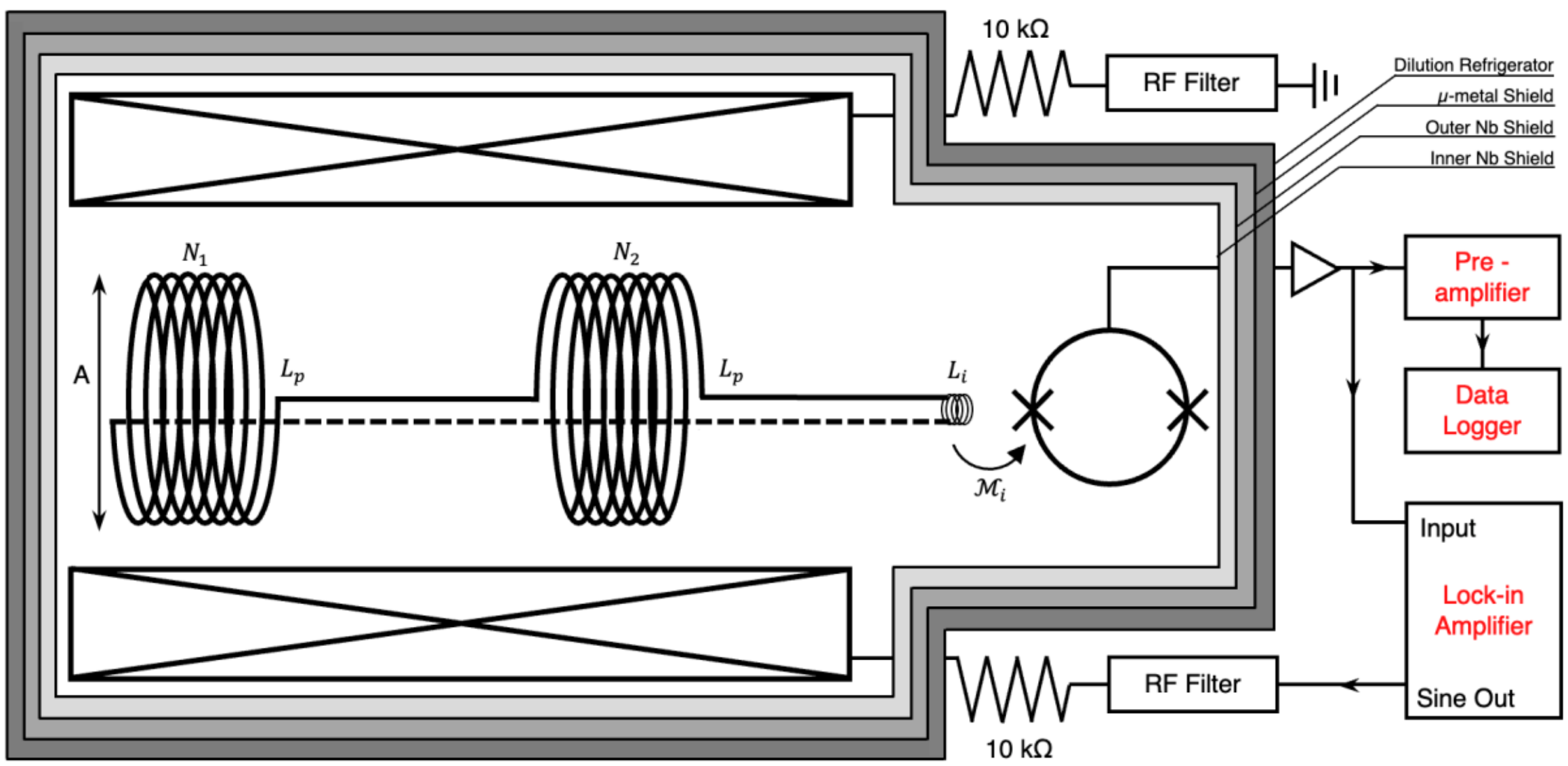

**Fig. S2:** The schematic of our combined monopole noise spectrometer and AC susceptometer. The circuit diagram illustrates the simultaneous monopole flux-noise and AC susceptibility measurement.

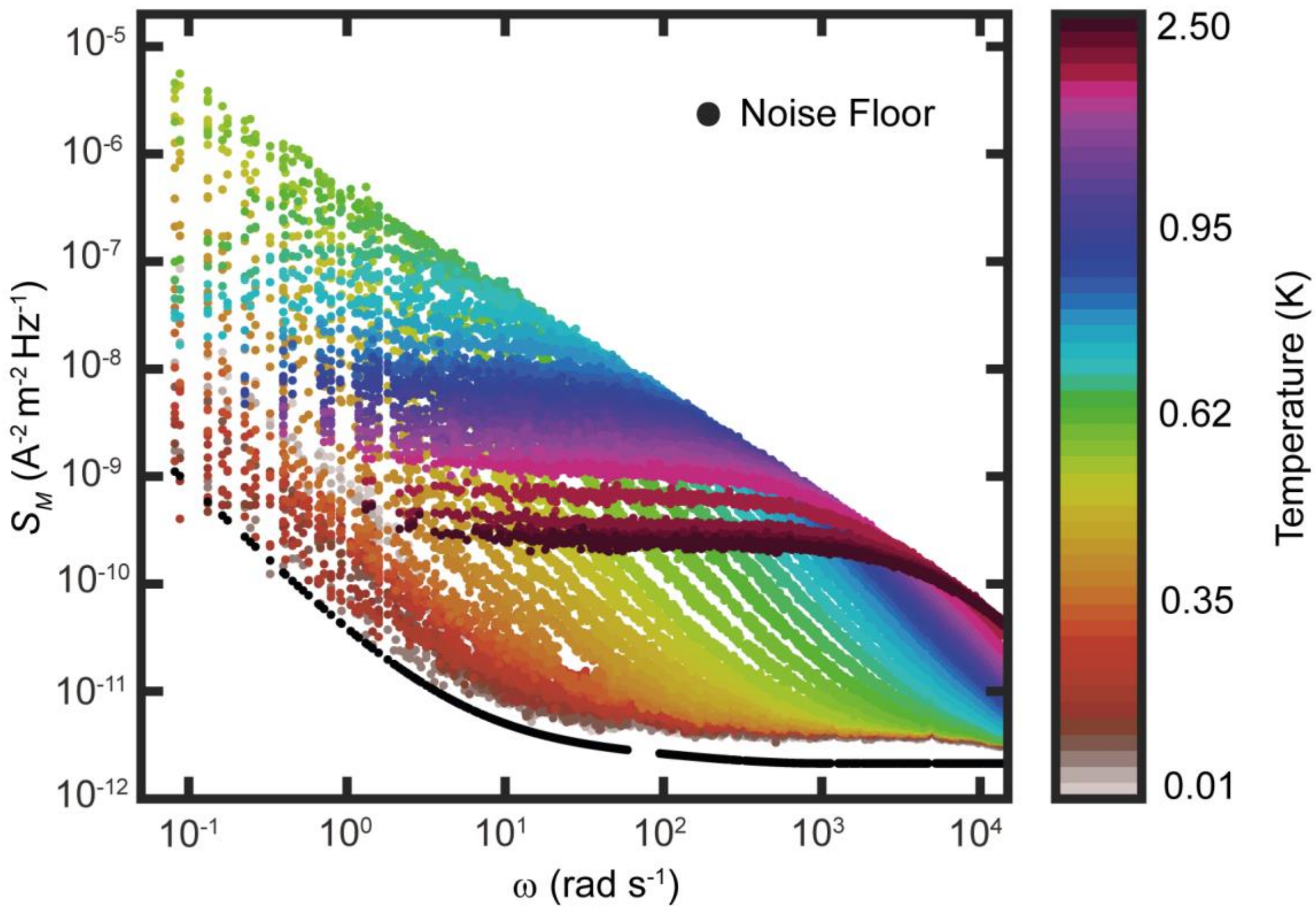

**Fig. S3:** Unprocessed magnetization noise power spectral density data $S_M(\omega, T)$ versus $T$. The measured empty-coil noise floor is plotted as a black curve and lies below the monopole noise spectra.

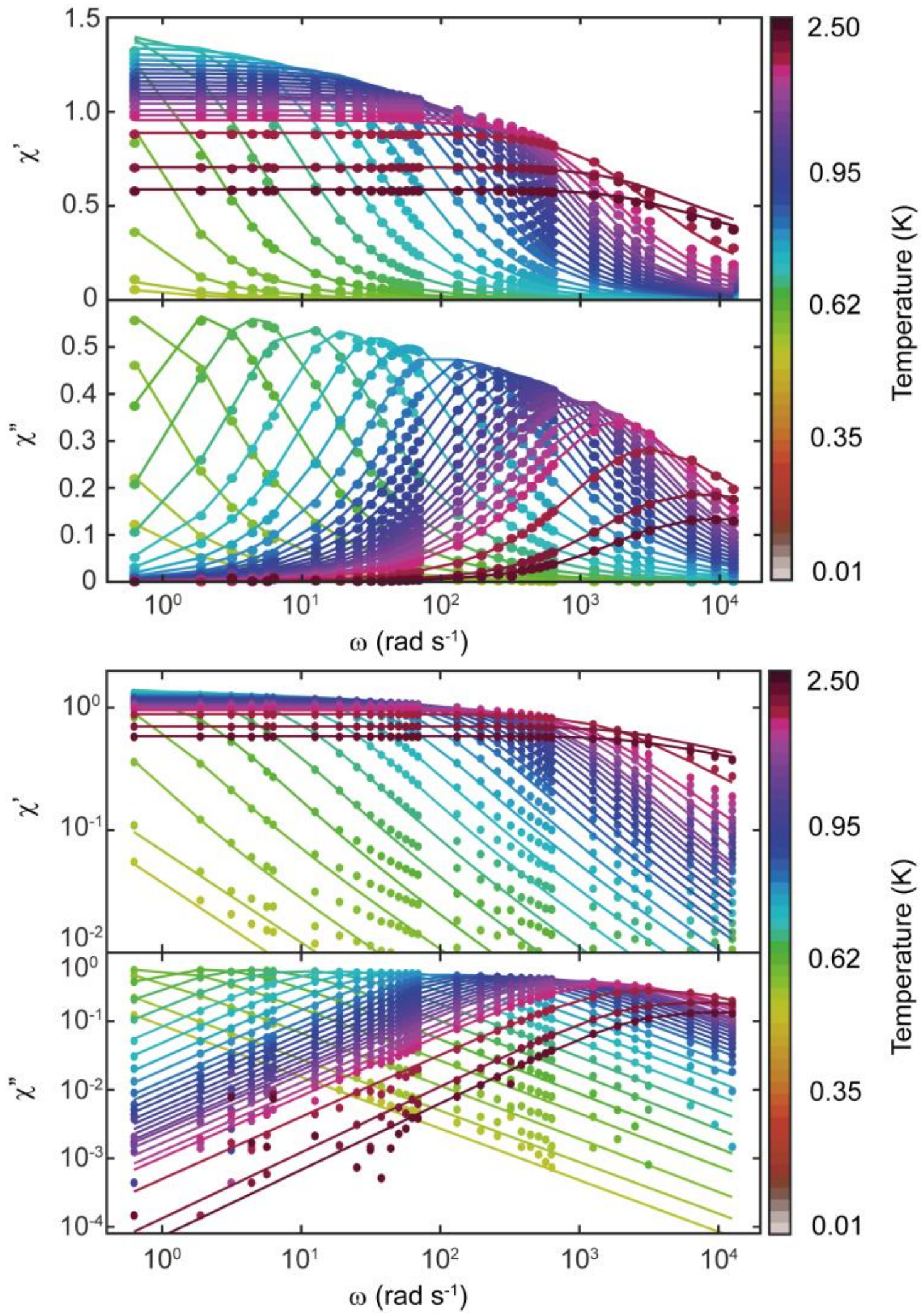

**Fig. S4:** The real component (top) of the magnetic AC susceptibility $\chi'(\omega, T)$ is fitted to its parametric equation S2. Below 500 mK the fit fails ($R^2 < 0.99$). The imaginary component (bottom) of the magnetic AC susceptibility $\chi''(\omega, T)$ is fitted to its parametric equation S3. The evolution of the monopole linear-response relaxation time is reflected clearly by the shift of the peak in $\chi''(\omega, T)$ towards lower frequencies as the temperature is decreased. Below 500 mK, where the peak is no longer in our experimental window, the fit fails ($R^2 < 0.99$). Data that cannot be parametrized by S3 are included in Fig. S5. The lower figure shows the identical susceptibility data and corresponding fits presented on a log-log scale.

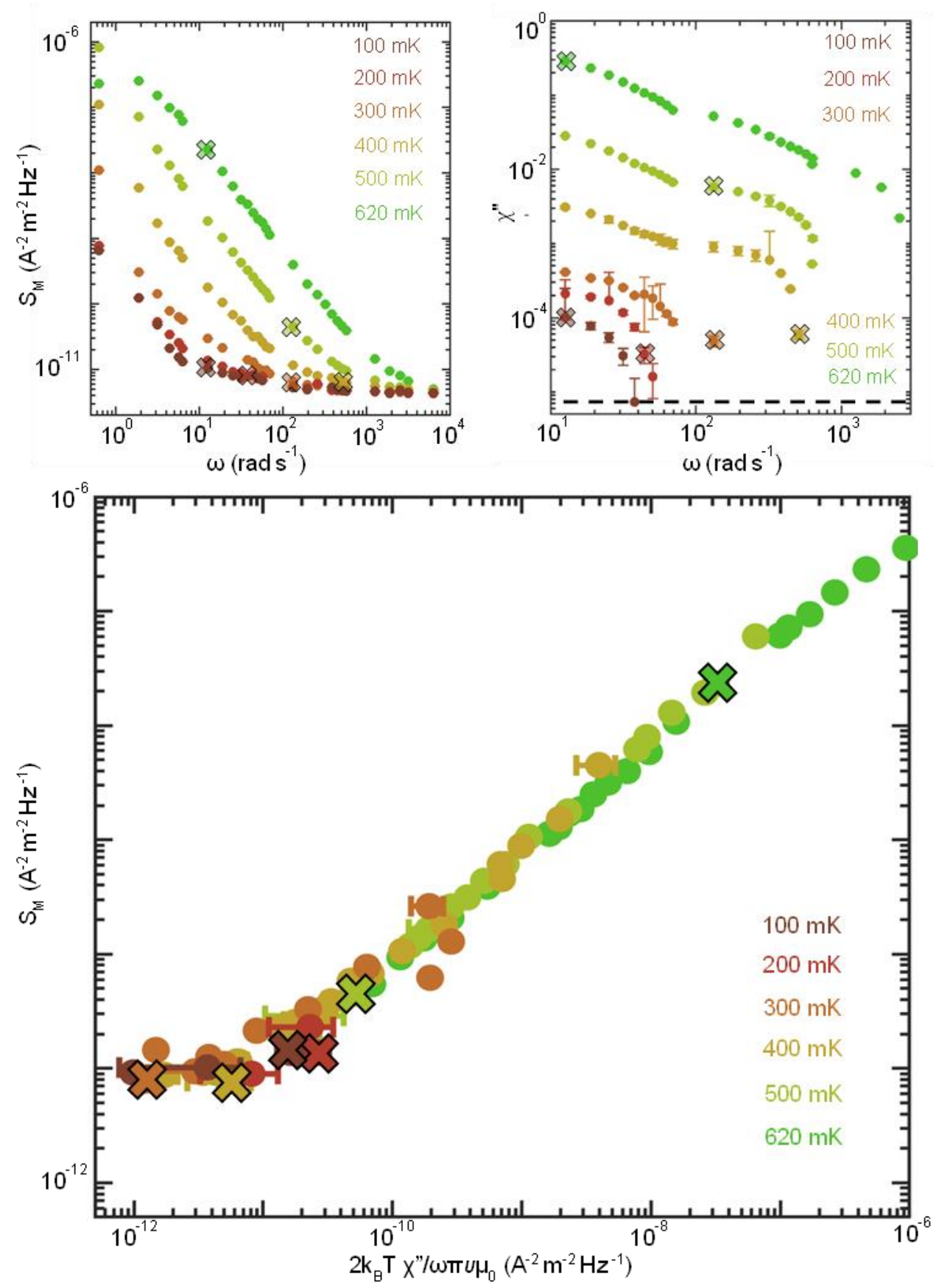

**Fig. S5:** Magnetization noise (top left) $S_M$ at 100 mK, 200 mK, 300 mK, 400 mK, 500 mK, and 620 mK. Each curve shows magnetization noise data at the corresponding frequencies to the susceptibility measurements in the next panel. The error in the noise is less than 1% of the signal in all cases, so the error bars are not included beyond this panel. Imaginary susceptibility (top right) χ" at the same temperatures as the top-left panel. The experimental noise floor is plotted at the base of the figure. Bottom: The left-hand side of the fluctuation-dissipation relation S11 (Y) is compared against the right-hand side (X). One data point at each temperature is represented by an 'X' as a guide to the eye. The same points are highlighted in the top panels to identify the pair of unprocessed noise and susceptibility values yielding that data point. At temperatures below 300 mK, a violation of the fluctuation-dissipation theorem is observed, as the linear relationship between the simultaneously measured magnetization noise and imaginary susceptibility fails.

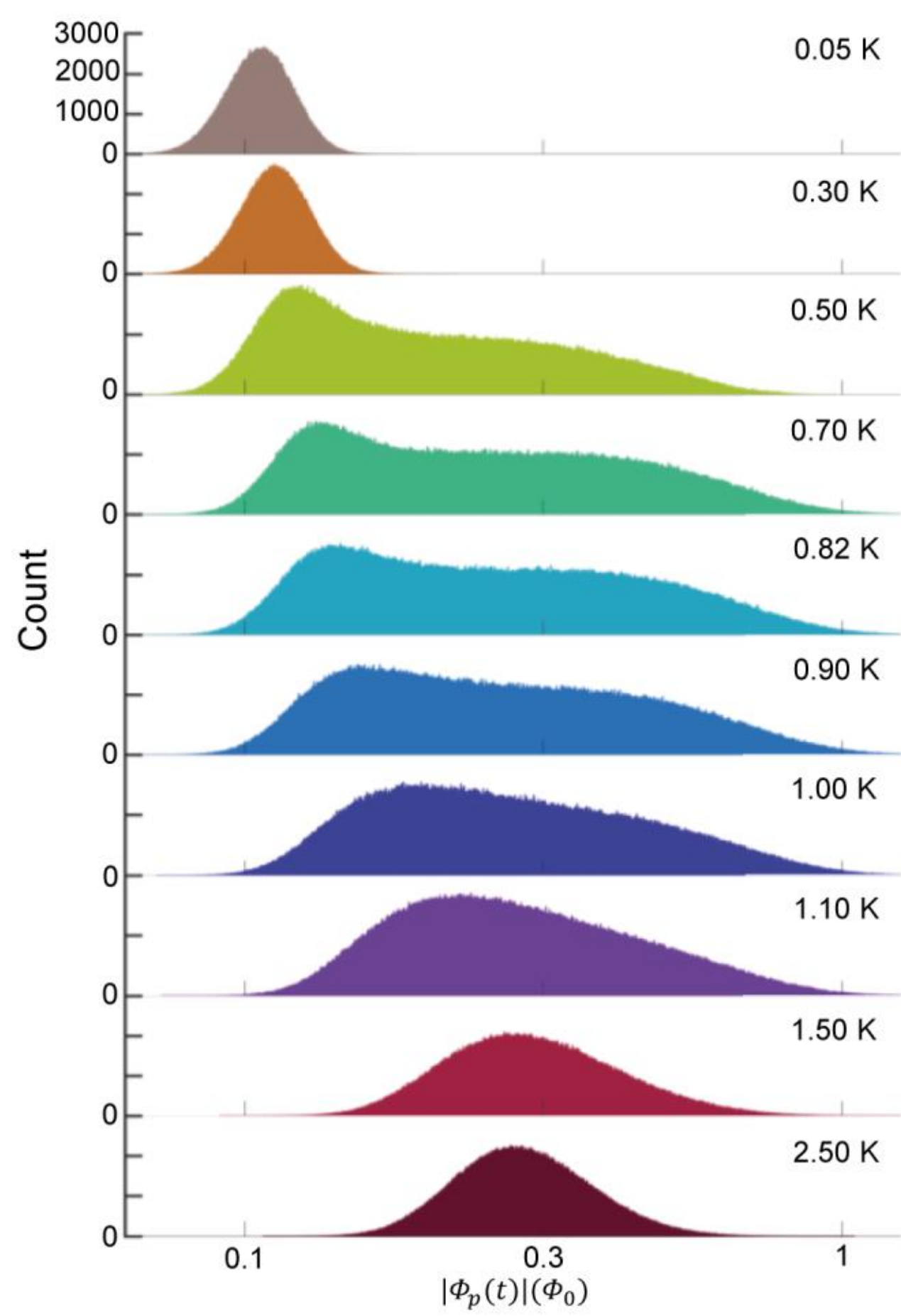

**Fig. S6:** Typical histograms of $|\Phi_p(t)|$. Conventional monopole current with a single Gaussian distribution persists until $T \approx 1500$ mK. A second current source, due to intense monopole current bursts appears below this temperature resulting in a bimodal distribution of probabilities. Below $T \lesssim 250$ mK the current bursts disappear.

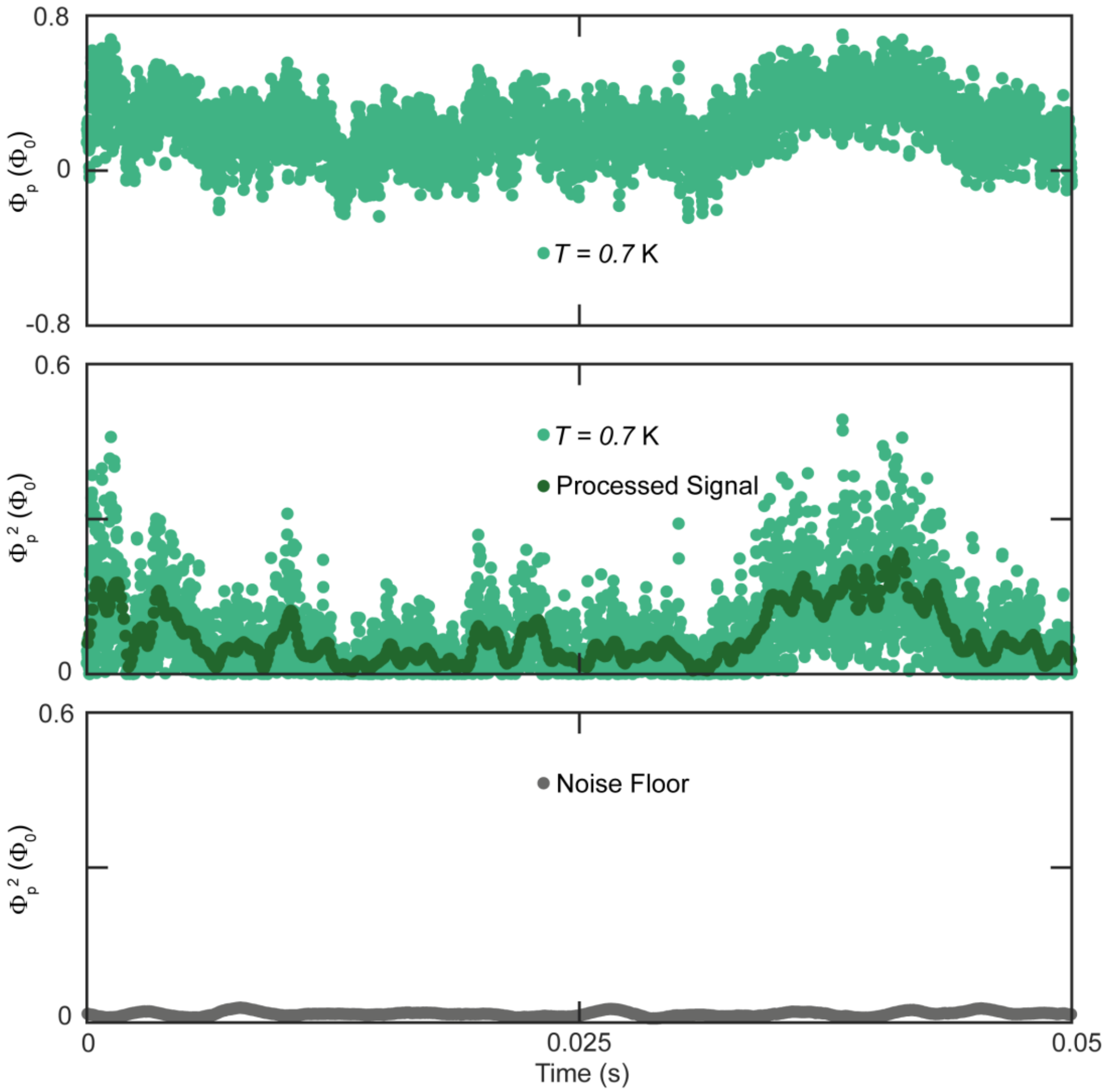

**Fig. S7:** Note that $\Phi_p^2$ and energy are considered equivalent here due to their linear relationship as described in Eqn. S16. (Top) A typical flux signal $\Phi_p$ measured at 700 mK. (Middle) The square of the flux signal $\Phi_p^2$ is calculated and the signal is then averaged in an 80 μs window. The averaged signal is layered on top of $\Phi_p^2$. (Bottom) The same routine is applied to the empty coil signal. The flux signal is considerably reduced in the empty coil data.

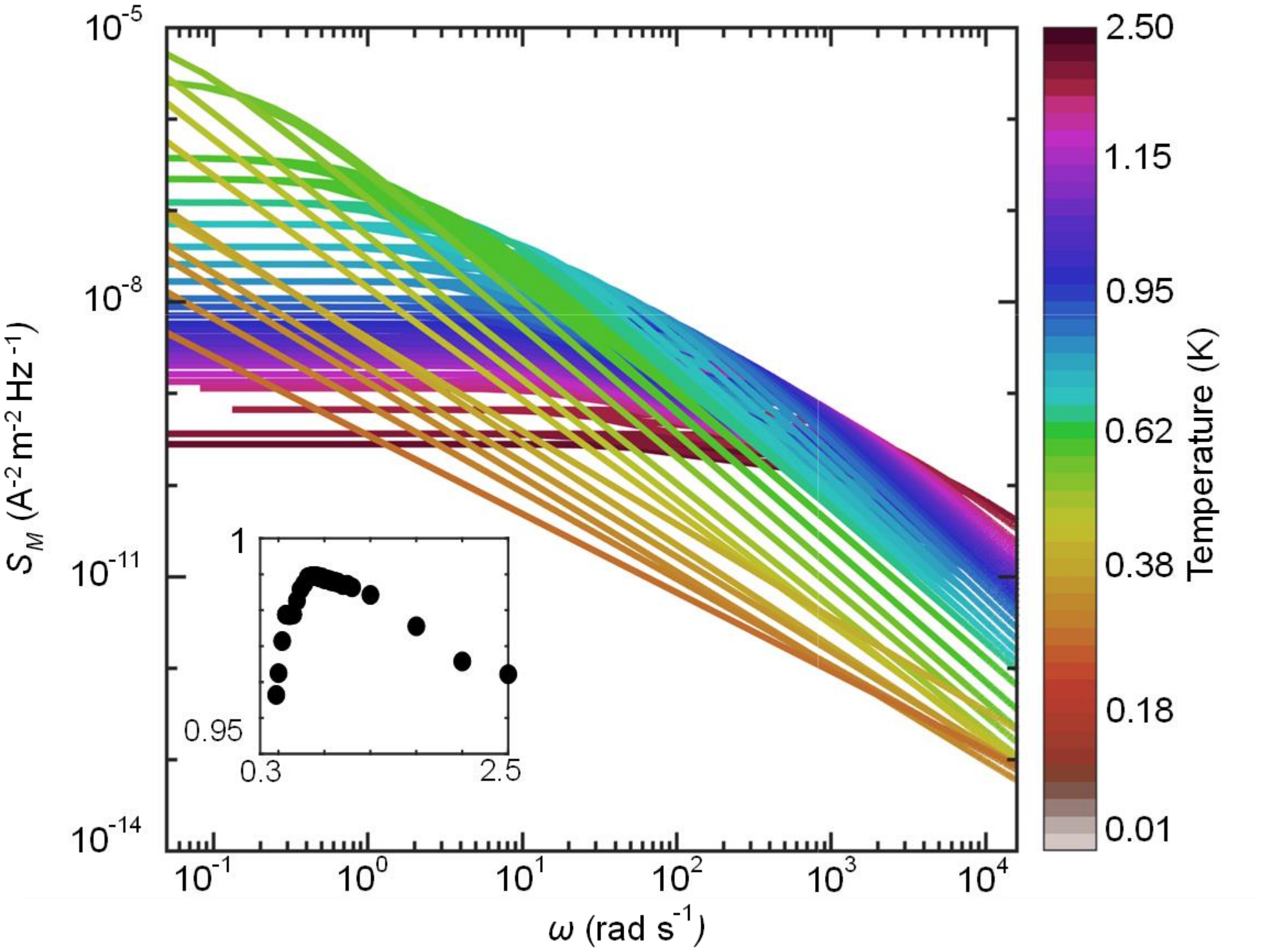

**Fig. S8:** Fitted magnetization noise power spectral density $S_M(\omega, T)$ data versus $T$. The noise is well described ($R^2 > 0.95$) by monopole generation/recombination above 300 mK. Below this temperature, fits are excluded.

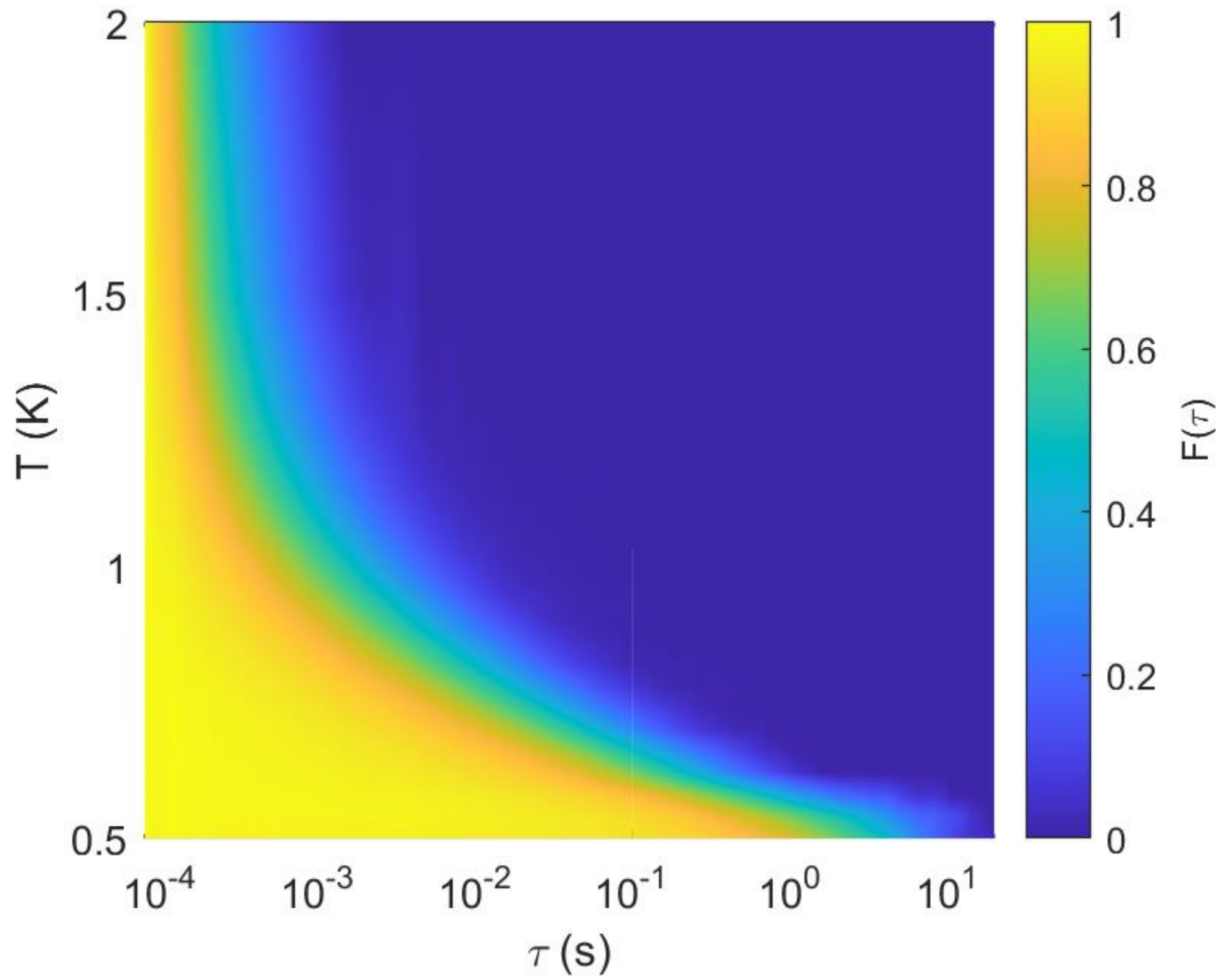

**Fig. S9:** Autocorrelation function data $F(\tau, T)$ shown as a function of lag time $\tau$ and temperature $T$. $F(\tau, T)$ shows the dramatically slowing dynamics of the monopoles with falling temperature. The increase in correlated monopole motion is evident as the monopole fluid enters the supercooled regime.

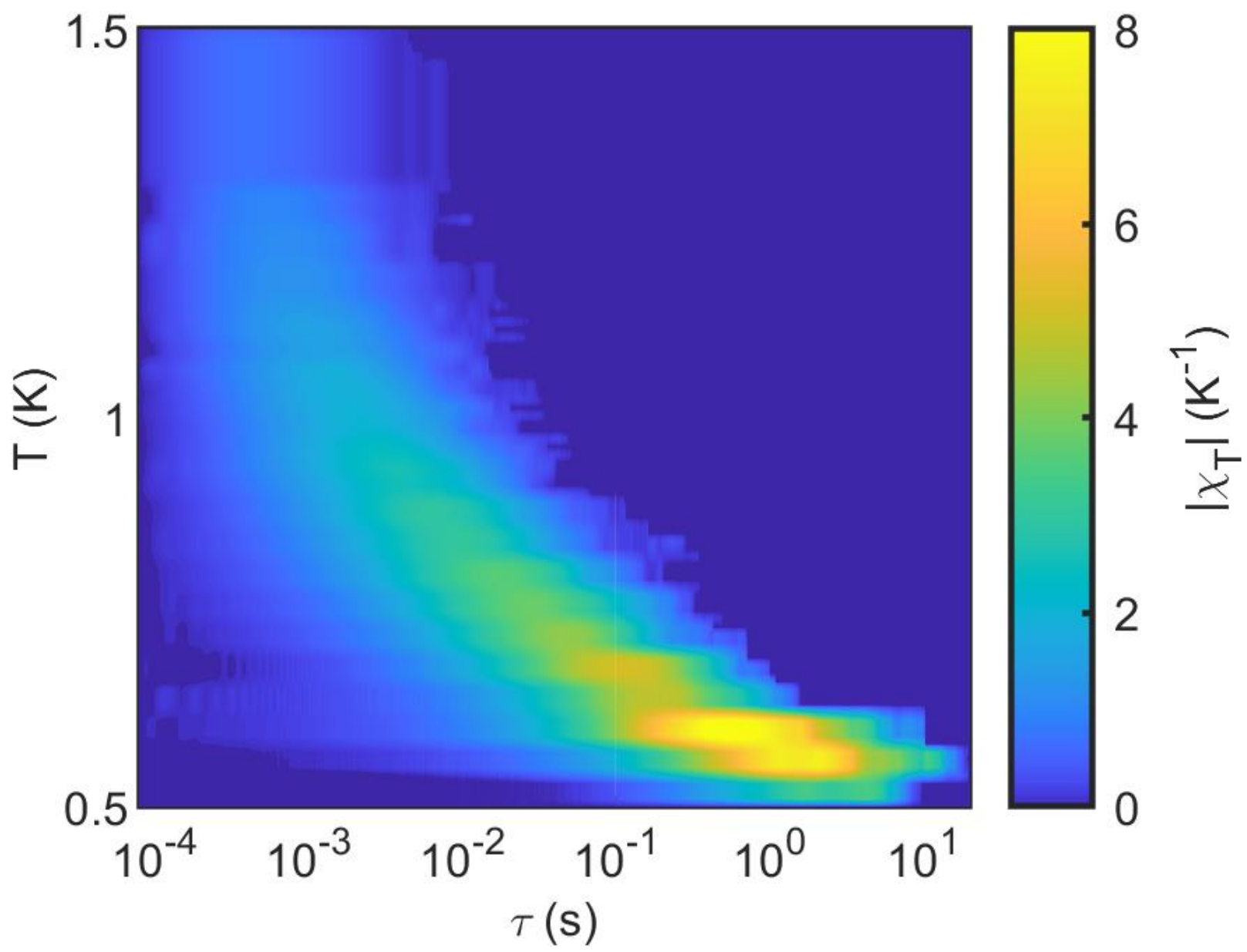

**Fig. S10:** Response function $\chi_T(\tau, T) = \partial F(\tau, T)/\partial T$ data shown as a function of lag time $\tau$ and temperature $T$, directly calculated from the data in Fig. S9.

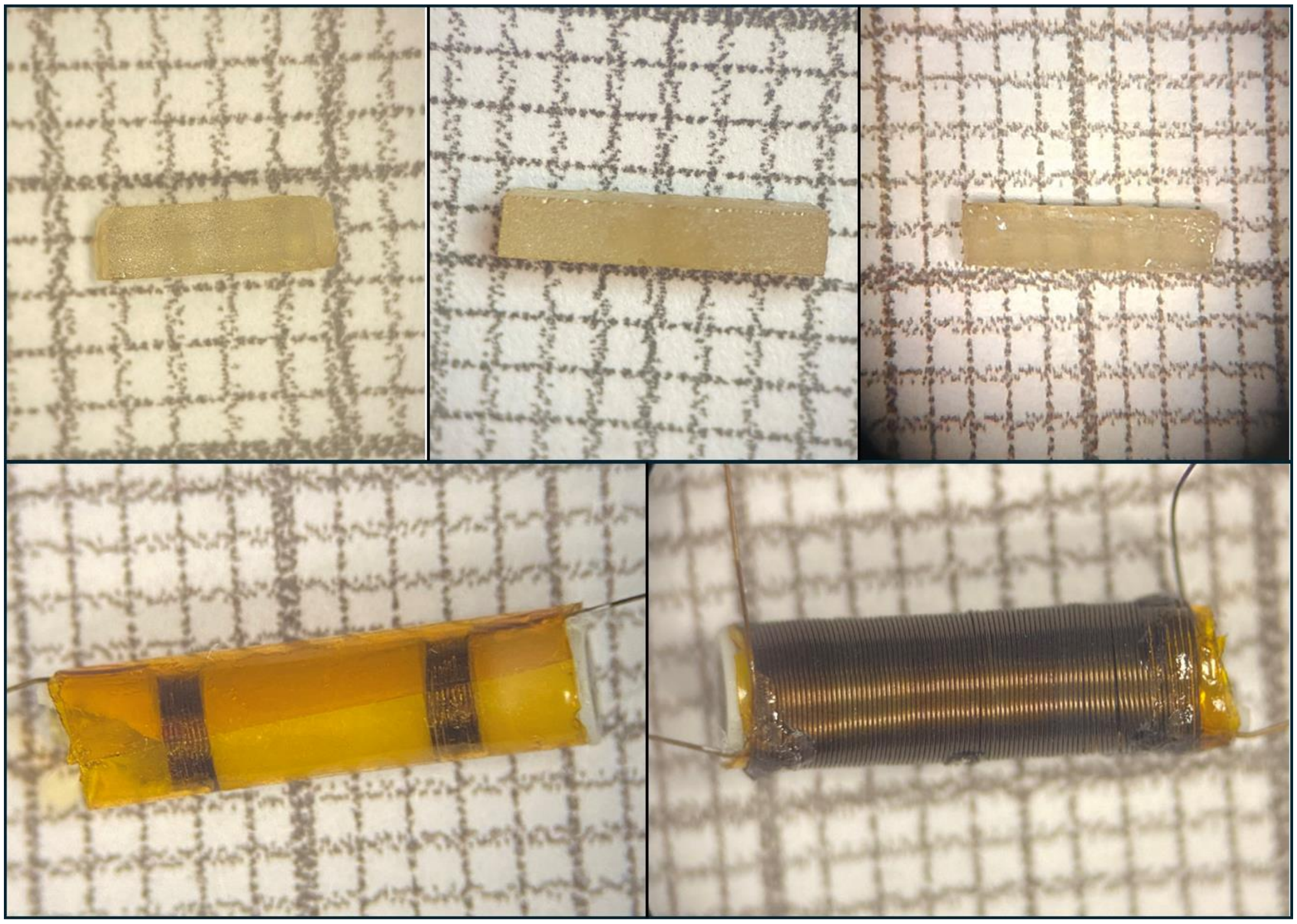

**Fig. S11:** Top Left: DTO Sample 1 with dimensions 0.74 mm x 1.48 mm x 4.97 mm. Top Middle: DTO Sample 2 with dimensions 0.98 mm x 1.31 mm x 6.23 mm. Top Right: DTO Sample 3 with dimensions 0.80 mm x 1.27 mm x 6.62 mm. Bottom Left: Pickup + Astatic coil assembly. The sample is inserted from the right into the pickup (rightmost) coil. Bottom Right: Drive coil wound around the pickup coil assembly. A layer of thin Kapton tape protects the inner wires while the drive coil is assembled.

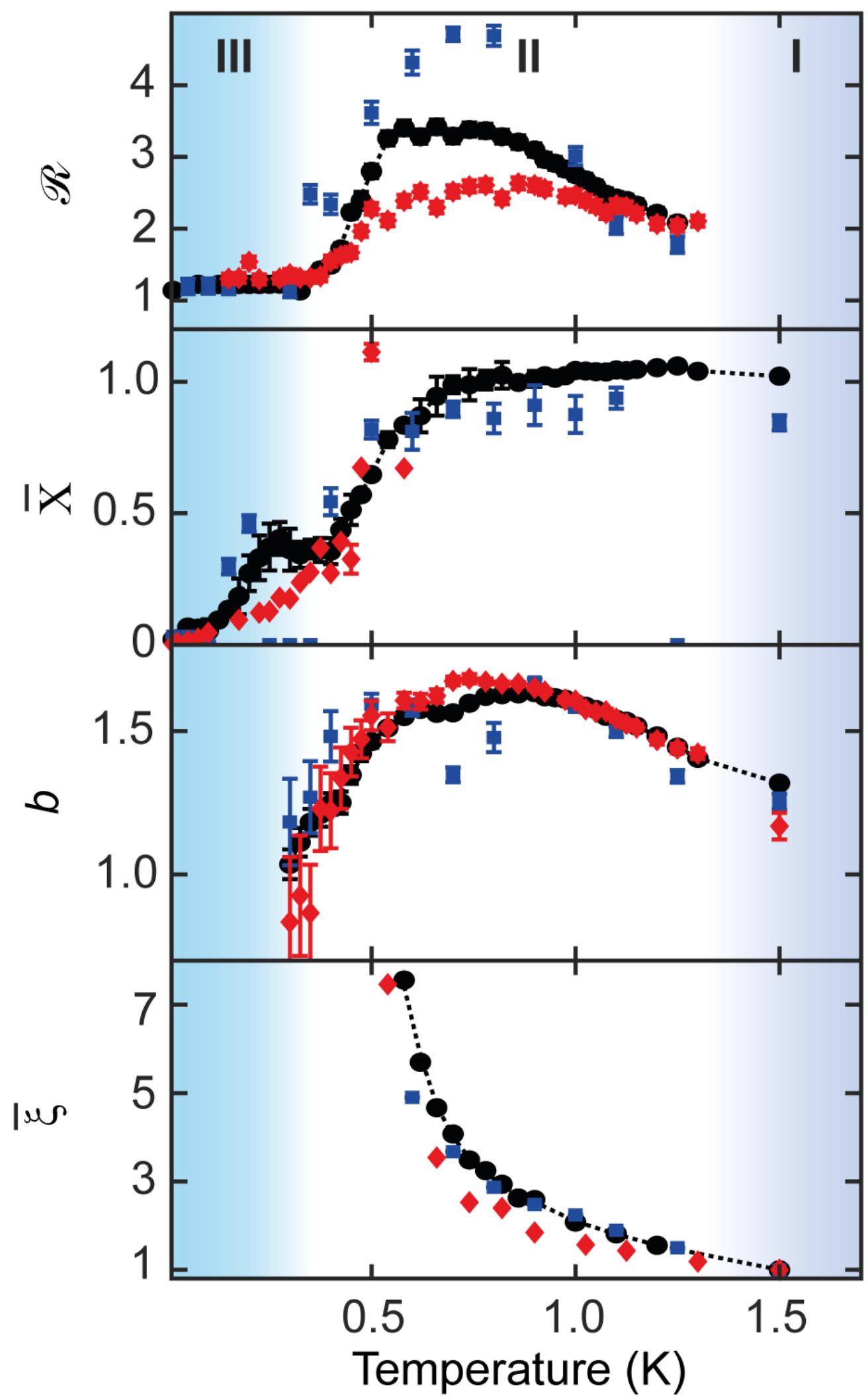

**Fig. S12:** Each sample studied in this work produced the same phenomenologies, demonstrating qualitative repeatability of the experiment. Sample 1 is shown as blue squares, Sample 2 is shown as black dots and Sample 3 is shown as red diamonds. Changes in magnitude of the noise can be attributed to geometric differences between samples.

**Movie & Audio S1:** Top: The evolution of the flux noise $\Phi_p(t, T)$ with falling temperature from $T = 2500$ mK to $T = 15$ mK. The flux noise signal, as it appears on screen, is converted to an audio signal, and played over the video. Bottom**:** The simultaneous evolution of the monopole noise and monopole current burst energies at temperatures $15$ mK $< T < 2500$ mK.

## Supplementary References

51 M. Udagawa, L. Jaubert, *Spin Ice* (Springer, 2021).

52 D. Pomaranski *et al.,* Absence of Pauling's residual entropy in thermally equilibrated $Dy_2Ti_2O_7$. *Nat. Phys*. **9**, 353-356 (2013).

53 C. Paulsen *et al*., Far-from-equilibrium monopole dynamics in spin ice. *Nat. Phys.* **10**, 135–139 (2014).

54 H. M. Revell *et al.,* Evidence of impurity and boundary effects on magnetic monopole dynamics in spin ice. *Nat. Phys*. **9**, 34-37 (2013).

55 J. A. Quilliam, I. R. Yaraskavitch, H. A. Dabkowska, B. D. Gaulin, J. B. Kycia, Dynamics of the magnetic susceptibility deep in the Coulomb phase of the dipolar spin ice material $Ho_2Ti_2O_7$. *Phys. Rev. B* **83**, 094424 (2011).

56 M. J. Jackson *et al*., Dynamic behavior of magnetic avalanches in the spin-ice compound $Dy_2Ti_2O_7$. *Phys. Rev. B* **90**, 064427 (2014).

57 C. Paulsen *et al.,* Nuclear spin assisted quantum tunnelling of magnetic monopoles in spin ice. *Nat Commun* **10**, 1509 (2019).

58 K. Matsuhira *et al.,* Spin dynamics at very low temperatures in spin ice $Dy_2Ti_2O_7$. *J. Phys. Soc. Jpn* **80**, 123711 (2011).

59 L. R. Yaraskavitch *et al*., Spin dynamics in the frozen state of the dipolar spin ice $Dy_2Ti_2O_7$. *Phys. Rev. B* **85**, 020410 (2012).

60 K. Matsuhira, Y. Hinatsu, T. Sakakibara, Novel dynamical magnetic properties in the spin ice compound $Dy_2Ti_2O_7$. *J. Phys.: Condens. Matter* **13**, L737 (2001).

61 D. J. P. Morris *et al.,* Dirac Strings and Magnetic Monopoles in the Spin Ice $Dy_2Ti_2O_7$. *Science* **326**, 411-414 (2009).

62 T. Stöter *et al.*, Extremely slow nonequilibrium monopole dynamics in classical spin ice. Phys. Rev. B 101, 224416 (2020).

63 J. Snyder *et al*., Low-temperature spin freezing in the $Dy_2Ti_2O_7$ spin ice. *Phys. Rev. B* **69**, 064414 (2004).

64 L. Bovo, J. A. Bloxsom, D. Prabhakaran, G. Aeppli, S. T. Bramwell, Brownian motion and quantum dynamics of magnetic monopoles in spin ice. *Nat. Commun* **4**, 1535-1542 (2013).

65 J. Snyder, J. S. Slusky, R. J. Cava, P. Schiffer, How 'spin ice' freezes. *Nature* **413**, 48-51 (2001).

66 J. Shi *et al*., Dynamical magnetic properties of the spin ice crystal $Dy_2Ti_2O_7$. *J. Magn. Magn. Mater*. **310,** 1322-1324 (2007).
67 R. A. Borzi *et al*. Intermediate magnetization state and competing orders in $Dy_2Ti_2O_7$ and $Ho_2Ti_2O_7$. *Nat. Commun* **7**, 12592 (2016).
68 P. K. Yadav, C. Upadhyay, Quantum criticality in geometrically frustrated $Ho_2Ti_2O_7$ and $Dy_2Ti_2O_7$ spin ices. *J. Magn. Magn. Mater.* **482**, 44-49 (2019).
69 H. Takatsu *et al*., Universal dynamics of magnetic monopoles in two-dimensional Kagomé ice. *J. Phys. Soc. Jpn* **90**, 123705 (2021).
70 A. B. Eyvazov *et al.,* Common glass-forming spin-liquid state in the pyrochlore magnets $Dy_2Ti_2O_7$ and $Ho_2Ti_2O_7$. *Phys. Rev. B* **98**, 214430 (2018).
71 S. Giblin, S. T. Bramwell, P. C. W. Holdsworth, D. Prabhakaran, I. Terry, Creation and measurement of long-lived magnetic monopole currents in spin ice. *Nat. Phys*. **7** 252-258 (2011).
72 M. Orendáč *et al.,* Magnetocaloric study of spin relaxation in dipolar spin ice. *Phys. Rev B* **75**, 104425 (2007).
73 C. Donati, S. Franz, S. C. Glotzer, G. Parisi, Theory of non-linear susceptibility and correlation length in glasses and liquids. *J. Non-Cryst. Solids* **307-310**, 215 (2002).
74 J. P. Hansen, I. R. Mc Donald, *Theory of Simple Liquids* (Elsevier, Amsterdam, 1986).
75 J. L. Lebowitz, J. K. Percus, L. Verlet, Ensemble Dependence of Fluctuations with Application to Machine Computations. *Phys. Rev.* **153,** 250 (1967).
76 G. Tarjus, D. Kivelson, S. Mossa, C, Alba-Simionesco, Disentangling density and temperature effects in the viscous slowing down of glassforming liquids*. J. Chem. Phys* **120,** 6135-6141 (2004).
77 L. Jaubert, P. C. W. Holdsworth, Signature of magnetic monopole and Dirac string dynamics in spin ice. *Nat. Phys.* **5**, 258–261 (2009).
78 K. Binder, A. P. Young, Spin. Glasses: Experimental facts, theoretical concepts, and open questions. *Rev. Mod. Phys*. **58**, 901 (1986).
79 C. Crauste-Thibierge *et al*., Evidence of Growing Spatial Correlations at the Glass Transition from Nonlinear Response Experiments. *Phys. Rev. Lett*. **104**, 165703 (2010).
80 Th. Bauer, P. Lunkenheimer, A. Loidl, Cooperativity and the Freezing of molecular Motion at the Glass Transition, *Phys. Rev. Lett*. **111**, 225702 (2013).

81 S. Albert *et al*., Fifth-order susceptibility unveils growth of thermodynamic amorphous order in glass-formers. *Science* **352**, 1308 (2016).

82 U. Tracht *et al*., Length Scale of Dynamic Heterogeneities at the Glass Transition Determined by Multidimensional Nuclear Magnetic Resonance. *Phys. Rev. Lett*. **81**, 2727 (1998).

83 A. Smith, *et al.,* The demagnetizing field of a nonuniform rectangular prism. *J. Appl. Phys.* **107**, 103910 (2010).

84 A. Ghasemi, A. Scheie, J. Kindervater, S. M. Koohpayeh, The pyrochlore Ho2Ti2O7: Synthesis, crystal growth, and stoichiometry. *J. Cryst. Growth* **500**, 38 (2018).

85 C. Castelnovo, R. Moessner, S. L. Sondhi, Debye-Hückel theory for spin ice at low temperature. *Phys. Rev. B* **84**, 144435 (2011).